\documentclass[aps,prb,twocolumn,amsmath,amssymb,showpacs,superscriptaddress]{revtex4-1}
\usepackage{graphicx}
\usepackage{dcolumn}
\usepackage[mathlines]{lineno}
\usepackage{hyperref}
\usepackage{bm}
\usepackage{epstopdf}
\usepackage{epsfig}
\usepackage{bbold}
\usepackage{ulem}
\usepackage{notes2bib}

\newcommand{\bra}[1]{\ensuremath{\left\langle#1\right|}}
\newcommand{\ket}[1]{\ensuremath{\left|#1\right\rangle}}

\newcommand{\brakket}[3]{\ensuremath{\langle{#1}|{#2}|{#3}\rangle}}

\usepackage{color}

\renewcommand{\vec}[1]{\mathbf{#1}}

\newcommand{\BN}[1]{{\color{black} #1}}

\begin{document}
	\title{Imaging \BN{and certifying high-dimensional entanglement} with a single-photon avalanche diode camera}
	\author{Bienvenu~Ndagano \normalfont\textsuperscript{1,\textasteriskcentered,$\dagger$} }
	\noaffiliation
	\author{Hugo Defienne,\normalfont\textsuperscript{1,\textasteriskcentered}} 
	\noaffiliation
	\author{Ashley Lyons}
	\affiliation{School of Physics and Astronomy, University of Glasgow, Glasgow G12 8QQ, UK}
	\author{Ilya Starshynov}
	\affiliation{School of Physics and Astronomy, University of Glasgow, Glasgow G12 8QQ, UK}
	\author{Federica Villa}
	\affiliation{Politecnico di Milano, Dipartimento di Elettronica, Informazione e Bioingegneria, Piazza Leonardo da Vinci 32, 20133, Milano, Italy}
	\author{Simone Tisa}
	\affiliation{Micro Photon Device SRL, Via Waltraud Gebert Deeg 3g, I-39100, Bolzano, Italy}
	\author{Daniele Faccio,\normalfont\textsuperscript{1,$\ddagger$}}
	\noaffiliation
	\date{\today}

	\begin{abstract}
Spatial correlations between two photons are the key resource in realising many quantum imaging schemes. Measurement of the bi-photon correlation map is typically performed using single-point scanning detectors or single-photon cameras based on CCD technology. However, both approaches are limited in speed due to the slow scanning and  the low frame-rate of CCD-based cameras, resulting in data acquisition times on the order of many hours. Here we employ a high frame rate, single photon avalanche diode (SPAD) camera, to measure the spatial joint probability distribution of a bi-photon state produced by spontaneous parametric down-conversion, with statistics taken over $10^7$ frames. \BN{Through violation of an Einstein-Podolsky-Rosen criterion by 227 sigmas, we confirm the presence of spatial entanglement between our photon pairs. Furthermore  we certify, in just 140 seconds, an entanglement dimensionality of 48.} Our work demonstrates the potential of SPAD cameras in the rapid characterisation of \BN{photonic entanglement}, leading the way towards \BN{ real-time} quantum imaging and quantum information processing.
	\end{abstract}
	
	\maketitle
	
	\section{Introduction}
	
	Individual single photon avalanche diodes (SPADs) have long been the workhorse of many quantum optics experiments~\cite{Migdall_book,Hadfield2009}. This is  due largely to their single photon level sensitivity, and also to the Geiger mode operation which allows for straightforward methods of single photon discrimination and counting, provided the detector operates in the photon-starved regime. Furthermore, precise timing electronics results in an impulse response function (IRF) that can be as short as 20 ps~\cite{Cova_avalanche_1996}, which is ideal for measuring temporal correlations between multiple photons whilst reducing the influence of background radiation and dark counts. These properties make SPADs one of the leading technologies for measuring photon-photon correlations and entanglement. 
	
	Arrays of SPADs, or SPAD cameras, fabricated with standard CMOS technology, have been produced in recent years and are now commercially available (e.g. from Photon Force, Micro Photon Devices). The maturity of the technology has enabled the production of compact arrays~\cite{Rochas2003,Rochas2003b}, as well as the reduction of the cost per device through bulk manufacturing processes~\cite{Bruschini2019}. Thus far, imaging devices based on SPADs have demonstrated their capabilities in fluorescence lifetime imaging~\cite{Li2010,Henderson2018,Bruschini2019}, LiDAR~\cite{Bronzi2014,Finlayson2018,Lindner2018,Henderson2019}, non-line-of-sight imaging~\cite{Gariepy2016,NLOS2020}, imaging through strongly scattering media~\cite{Lyons2019b} \BN{and time-resolved correlation measurements~\cite{Lubin2019,Unternahrer2016}}. However, SPAD cameras have yet to make their mark due to their overall efficiency and resolution; the fill-factor of the earliest available SPAD cameras was on the order of a few percent~\cite{Gariepy2015,Bruschini2019} which, despite the quantum efficiency of the single SPAD pixels being on par with commercial single element SPADs, equates to a large overall loss. This high loss is particularly detrimental to the detection of quantum states formed of multiple photons as it scales with the power of the photon number. 
	
	CCD based single photon cameras have therefore been typically preferred for quantum imaging applications, where one of the routine tasks involves measuring spatial correlations between photons to build an image. This in turn is a result of the photon source of choice that is usually spontaneous parametric down-conversion (SPDC) in non-linear crystals, where a pump photon is converted with a given probability, into a pair photons. The governing law for this process is momentum conservation that ensures correlations between the photons in the pair~\cite{Jost1998, Abouraddy2001}. These correlations can be exploited for ghost imaging~\cite{Pittman1995,Moreau2019}, imaging at enhanced spatial resolution~\cite{Xu2015,Toninelli2019}, \BN{quantum-enhanced target detection~\cite{Zhang2020} } and to distil an image  encoded in quantum states in the presence of classical background radiation~\cite{Defienne2019, Gregory2019}. \BN{We note that recently, Ianzano \textit{et al.} have demonstrated the measurement of polarisation entanglement using a camera~\cite{Ianzano2020}, owing to its high-temporal resolution (1.5 ns). However, this demonstration did not take full advantage of the spatial resolution of the camera to measure the spatial correlations in the photon pairs; these were spatially filtered using single-mode fibres before being measured with the camera. While this ensured mode-matching and spatial indistinguishability that are necessary for polarisation entanglement, the approach comes at the cost of reducing the number of modes to only one, insufficient for imaging, and limits the dimensionality of entanglement to two. }
	
	A quantity of particular interest in quantum imaging is the spatial bi-photon joint probability distribution (JPD) describing the correlations between photon pairs. The reconstruction of the JPD can be achieved through statistical averaging over a large number of intensity images~\cite{reichert_massively_2018} -- typically on the order of $10^6$ to $10^7$ images -- of identically prepared photon pairs. Given that CCD-based detectors provide frame rates on the order of $100$ frames/s, the total acquisition time can vary from a few hours to over a day. Such long acquisition times constitute a hindrance to the widespread adoption of quantum imaging schemes for practical applications.
	
	In the following, we show the reconstruction of the JPD in both position and momentum using a commercially available SPAD camera (SPC3, Micro Photon Devices). Correlation measurements in position and momentum space are averaged over a total of $10^7$ images, amounting to an acquisition time of 140 s. These measurements are of a quality that allows to demonstrate spatial entanglement through violation of an Einstein-Podolsky-Rosen (EPR) criterion~\cite{Einstein1935, howell_realization_2004} with a confidence of $227$ sigmas. We show an experimental study into the confidence of the violation as a function of the number of individual single-bit frames used for the calculation, highlighting the benefit of the high-frame rate enabled by the SPAD camera for fast characterisation of entanglement correlations. \BN{From our measurements in complementary position and momentum bases, we certify high-dimensional entanglement in up to $48$ dimensions. To this end, we used a robust method of entanglement certification developed by Bavaresco \textit{et al.} to compute the lower bound of the fidelity of the measured state with respect to a maximally entangled one in order to estimate its dimensionality~\cite{Bavaresco2018}}. The short acquisition times required to reconstruct the bi-photon JPD \BN{and certify high-dimensional entanglement} pave the way for the implementation of quantum imaging \BN{and information processing} schemes in real-time.  Indeed, we note that faster sensors, with frame rates up to 800 kframes/s, have been demonstrated~\cite{Gasparini2018} and that this work represents only the tip of the iceberg for the potential use of SPAD cameras within the field of quantum optics. 
	
	\section{Results}
\begin{figure}[htbp]
		\centering\includegraphics[width=1 \columnwidth]{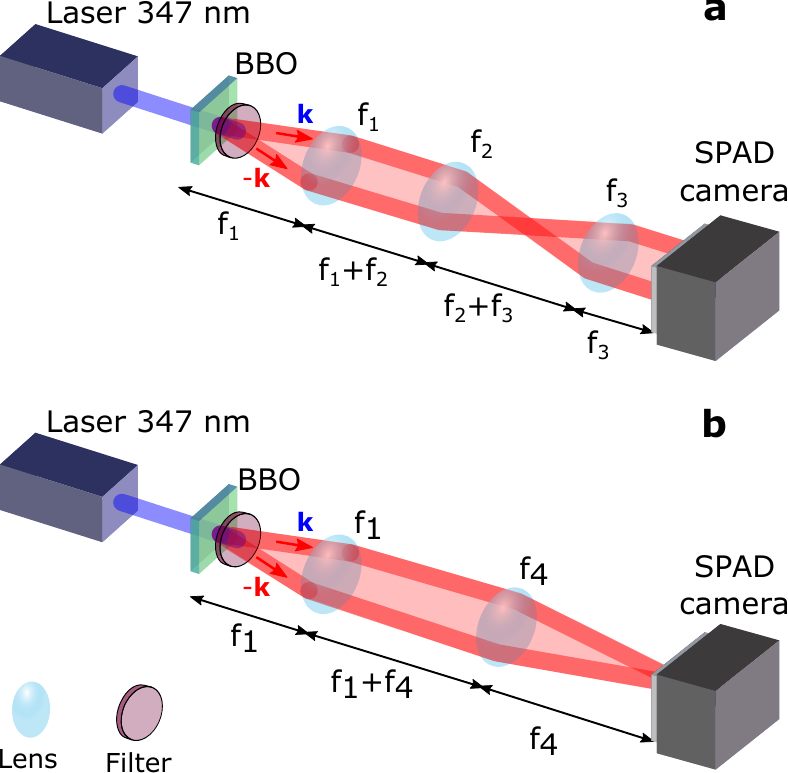}
		\caption{\label{Figure1} \textbf{Experimental scheme.} Spatially-entangled photon pairs are produced by spontaneous parametric down-conversion (SPDC) in a $\beta$-barium borate (BBO) using a $347$~nm pulsed pump laser with a repetition rate of 100~MHz. Spectral filters (SF) select near-degenerate photon pairs at $694 \pm 5$~nm. \textbf{a,} A three-lens system composed of $f_1=35$~mm, $f_2=100$~mm and $f_3=200$~mm, maps photon momenta onto pixels of a single-photon avalanche diode (SPAD) camera by Fourier imaging the crystal. This configuration is named FF-configuration.  \textbf{b,} position correlations are measured with two lenses, $f_1=35$~mm and $f_4=300$~mm, image the output surface of the BBO crystal onto the SPAD camera. This configuration is named NF-configuration.}
	\end{figure}

	\noindent\textbf{} Figure~\ref{Figure1} shows the experimental apparatus used to measure spatial correlations between photon pairs. Spatially entangled photon pairs are produced via spontaneous parametric down-conversion (SPDC) in a 0.5 mm long $\beta$-barium borate (BBO) crystal, cut for type-I phase matching. The crystal is pumped by a $347$ nm pulsed laser with a repetition rate of 100 MHz (pulse length of 10 ns), an average power of $50$ mW and is spatially filtered and collimated (not shown) to a beam diameter of $0.7$ mm ($1/e^2$). Spectral filters block the pump beam and select near-degenerate photon pairs at $694 \pm 5$ nm. \BN{Photon pairs are detected using an MPD-SPC3 single-photon avalanche diode (SPAD) camera with an array of $32 \times 64$ pixels, a fill factor of 80\% and a 150 $\mu$m pixel pitch. The quantum efficiency at 694 nm is around 9\%  and the dark count rate is 0.14 count per pixel per second. The SPAD camera has a nominal speed of $96$ kframes/s and is operated at its minimum exposure time (10 ns) and dead-time (50 ns). Our photon flux over the full array was 22.5 mega-counts per second}. \BN{To measure momentum correlations, we used a $4f$-telescope to image the Fourier plane of the BBO crystal surface onto the SPAD camera (FF-configuration), as shown in Fig.~\ref{Figure1}a. Figure~\ref{Figure1}b shows the configuration used to measure position correlations of photons pairs at the surface of the BBO crystal (NF-configuration); the output surface of the crystal is imaged onto the SPAD camera using a $4f$-telescope. } 
	
	\begin{figure}[htbp]
		\centering\includegraphics[width=1 \columnwidth]{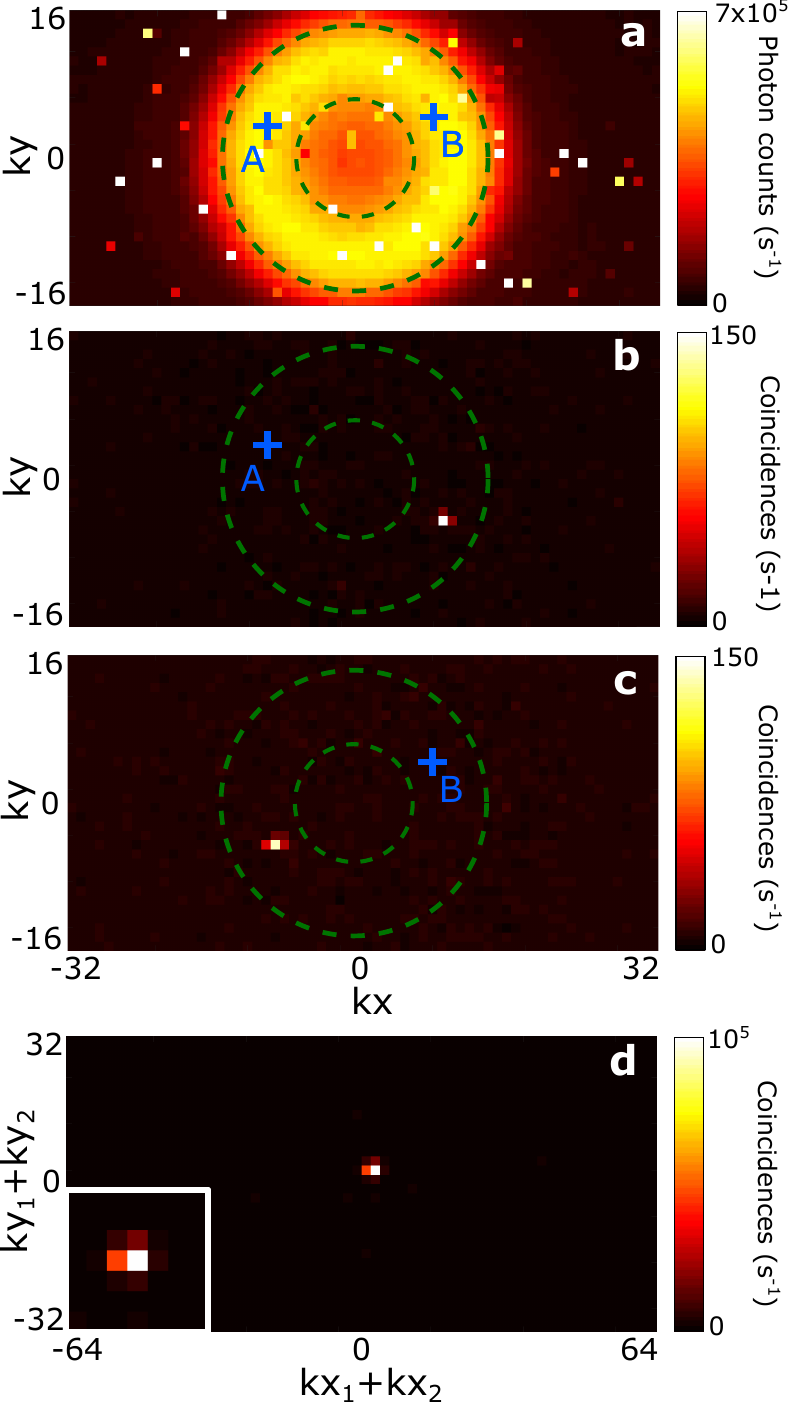}
		\caption{\label{Figure2} \textbf{Measurement of momentum correlations.} \textbf{a,} Intensity distribution of the SPDC light measured in the FF-configuration. \textbf{b} and \textbf{c}, Conditional probability distributions $\Gamma(\vec{k}|\vec{A})$ and  $\Gamma(\vec{k}|\vec{B})$ relative to two arbitrarily chosen positions $\vec{A}$ and $\vec{B}$ on the sensor, respectively. We measured an SNR of 320 (\textbf{b}) and 258 (\textbf{c}). \textbf{d}, Projection of the joint probability distribution (JPD) along the sum coordinates $\vec{k_1}+\vec{k_2}$. A measured momentum correlation width of $\Delta\vec{k} = 1.0666(7)\times 10^{-3} $ rad.$\mu\text{m}^{-1}$ is obtained using a Gaussian fit (see Methods). Spatial coordinates are in pixels and the analysis was performed on a total of $10^7$ images.}
	\end{figure}
	
	\begin{figure}[htbp]
		\centering\includegraphics[width=1 \columnwidth]{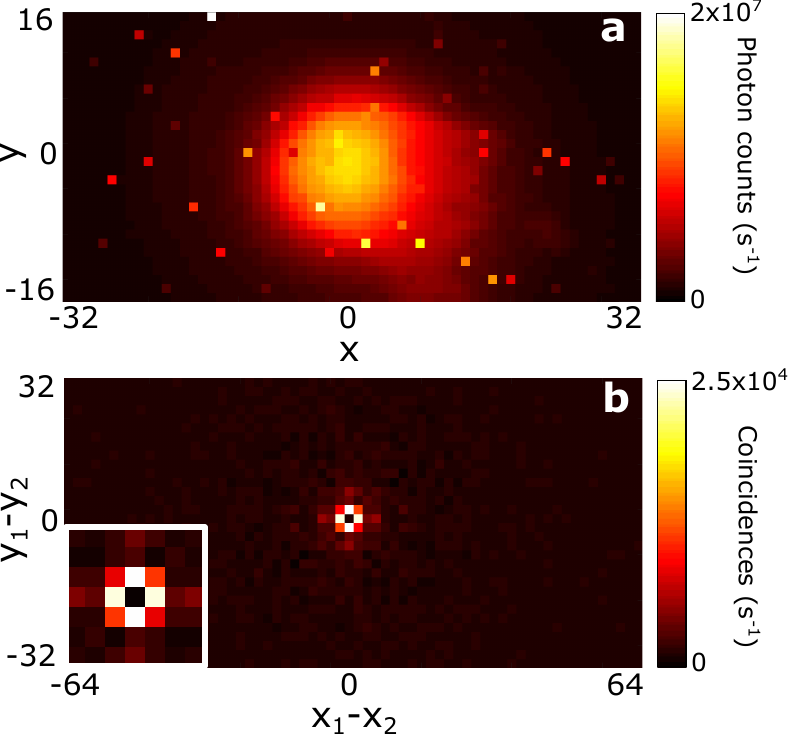}
		\caption{\label{Figure3} \textbf{Measurement of position correlations.} \textbf{a,} Intensity distribution of the SPDC light in the NF-configuration. \textbf{b,} Projection of joint probability distribution along the minus-coordinates $\vec{r_1}-\vec{r_2}$. A measured position correlation width $\Delta\vec{r} = 4.3(2)$ $\mu\text{m}$ was obtained using a Gaussian fit (see Methods). Spatial coordinates are in pixels and the analysis was performed on a total of $10^7$ images.}
	\end{figure} 
	
	\ \\
	\noindent\textbf{Measuring EPR entanglement.} We demonstrate the presence of spatial entanglement in SPDC light by violating the Einstein-Podolsky-Rosen (EPR) criterion~\cite{howell_realization_2004}
	\begin{equation}
	\Delta\vec{r}\cdot\Delta\vec{k} > \frac{1}{2},
	\label{eq: EPR}
	\end{equation}
	where $\Delta\vec{r} = \Delta(\vec{r_1}-\vec{r_2})$ and $\Delta\vec{k} = \Delta(\vec{k_1}+\vec{k_2})$ are, respectively, measures of the correlation strength in position and momentum for a pair of photons labelled $1$ and $2$. Violation of a related EPR criterion has been achieved previously both in the case of scanning single-point detectors~\cite{howell_realization_2004} and by full-field imaging using an electron-multiplying CCD (EMCCD) cameras~\cite{moreau_realization_2012, Edgar2012}.
	
	To determine the strength of the transverse position and momentum correlations, we measure the spatial JPD of photon pairs in the two configurations described in Fig.~\ref{Figure1} using the SPAD camera. For a given pair of pixels located at positions $\vec{r_i}$ and $\vec{r_j}$ of the sensor, an element of the JPD denoted $\Gamma(\vec{r_i},\vec{r_j})$ represents the joint probability of detecting photon $\vec{i}$ of a pair at pixel $\vec{r_i}$ and photon $\vec{j}$ at pixel $\vec{r_j}$. $\Gamma(\vec{r_i},\vec{r_j})$ is calculated from a set of $N$ frames using the formula~\cite{defienne_general_2018-2}
	\begin{equation}	
	\label{eq: JPD intensity}
	\Gamma(\vec{r_i},\vec{r_j}) = \frac{1}{N} \sum_{l=1}^N I_l(\vec{r_i})I_l(\vec{r_j}) - \frac{1}{N^2} \sum_{m,n=1}^{N} I_m(\vec{r_i})I_{n}(\vec{r_j}), 
	\end{equation}
	where $I_l(\vec{r_i}) \in \{0,1\}$ is a binary value returned by the SPAD camera at pixel $\vec{r_i}$ in the $l^{th}$ frame. The first term is an average value of the coincidence detection of photons belonging to either the same entangled pair (genuine coincidence) or different entangled pairs (accidental coincidence). Since multiple photon pairs can be detected during the time of an exposure, the contribution of accidental coincidences is generally greater than the genuine ones. The second term is an average value of accidental coincidences. Therefore, a subtraction between these two terms leaves only an average value of genuine coincidences that is $\Gamma(\vec{r_i},\vec{r_j})$ (see Methods). \BN{Note that, while subtracting the accidental coincidences is important for our approach to work, this process also has the drawback of increasing the noise in the reconstructed JPD.}
	
	\ \\
	\noindent\textbf{Reconstruction of bi-photon correlations.} Figure~\ref{Figure2} shows results of JPD measurements performed in the FF-configuration to study momentum correlations. Figure~\ref{Figure2}a shows the direct intensity image reconstructed from the sum of $10^7$ frames, which corresponds to a total acquisition time of 140 s. This image represents the probability of detecting a photon with a given momentum $\vec{k}=(k_x,k_y)$, with no information about the relative position of photons within the same pair (i.e. it is the marginal probability distribution). The JPD $\Gamma(\vec{k_1},\vec{k_2})$ is computed from this set of frames using Eq.~\ref{eq: JPD intensity} and is visualized using conditional projections. For example, Fig.s~\ref{Figure2}b and \ref{Figure2}c show the conditional spatial distributions $\Gamma(\vec{k}|\vec{A})$ and $\Gamma(\vec{k}|\vec{B})$ of two photons measured by the sensor in coincidence  and with high signal-to-noise ratio (SNR), with one-photon detected at arbitrarily chosen positions $\vec{A}$ and $\vec{B}$, respectively. 
	
	Due to momentum conservation in the SPDC process, the signal and idler photons are measured at $\pi$ radians in the transverse plane at the center of the marginal distribution. The centrosymmetry of momentum correlations is characterised by the presence of an intense peak of coincidences in the sum-coordinate projection of the JPD shown in Fig.~\ref{Figure2}d. The height of the peak corresponds to the sum of coincidences measured in all pairs of symmetric pixels, while its width gives the strength of the correlation, i.e., the momentum correlation width $\Delta\vec{k}$. Accounting for the effective magnification of our optical system, we measured $\Delta\vec{k} = 1.0666(7)\times 10^{-2} $ rad.$\mu\text{m}^{-1}$ using a Gaussian fitting model. \BN{This model follows from the double-Gaussian approximation applied to the two-photon wavefunction of photon pairs produced by SPDC in a thin crystal~\cite{Fedorov2009,Schneeloch2016} used in many experimental studies~\cite{moreau_realization_2012, Edgar2012,howell_realization_2004,reichert_massively_2018} (see Methods).}
	
	We repeated the above analysis in the NF-configuration shown in Fig.~\ref{Figure1}b to extract the position correlation width $\Delta\vec{r}$. The photons in entangled pairs are position correlated, i.e., they are born at the same position in the crystal and are expected to arrive at the SPAD sensor at the same position.  Figures~\ref{Figure3}a and b show, respectively, the direct intensity image and the projection of the JPD along the minus-coordinate reconstructed from a set of $10^7$ frames. A  coincidence peak is observed at the center of the minus-coordinate projection that demonstrates strong position correlation between photon pairs. The central pixel in the minus-coordinate projection has been set to zero because the SPAD camera does not resolve the number of photons detected per pixel and therefore cannot measure photon coincidences at the same pixel. Accounting for the optical magnification, we measured a position correlation width $\Delta\vec{r} = 4.3(2)\ \mu$m by fitting with a Gaussian model (see Methods). 
	
	The measured values of transverse position and momentum correlations width violate the EPR criterion in Eq.~(\ref{eq: EPR}): $\Delta\vec{r}\cdot \Delta\vec{k} = 4.6(2)\times10^{-2}<1/2$, thus demonstrating the presence of spatial entanglement. This violation has a confidence level of $C=227$ using the following definition:
	\begin{equation}
	\label{equconfidence}
	C=\frac{|1/2-\Delta\vec{r}\cdot \Delta\vec{k} |}{\sigma}
	\end{equation}
	where $\sigma = 10^{-3}$ is the uncertainty on the product $\Delta\vec{r}\cdot \Delta\vec{k}$. 
	
	\BN{In both position and momentum space, we observe that the correlation width is at most, as large as a single pixel on the SPAD camera. On one hand, this is due to the relatively large 150 $\mu$m pixel pitch that is for comparison,  over 10x larger than that of the EMCCD iXon 888 from Andor technologies. On the other hand, The magnification of our optical system was not large enough to  spread the correlation width over more pixels. The implication is that the true correlation width as given by the crystal is necessarily smaller than what we measured (see Methods). However, a more accurate and precise measurement of the correlation width will only lead to a tighter EPR violation. Note that this can now be realised with state-of-the-art SPAD sensors that boast megapixel-arrays and 2.2 $\mu$m pixel pitch \cite{Morimoto2020_1,Morimoto2020_2}}
	
	\ \\
		\begin{figure}[t]
		\centering\includegraphics[width=1 \columnwidth]{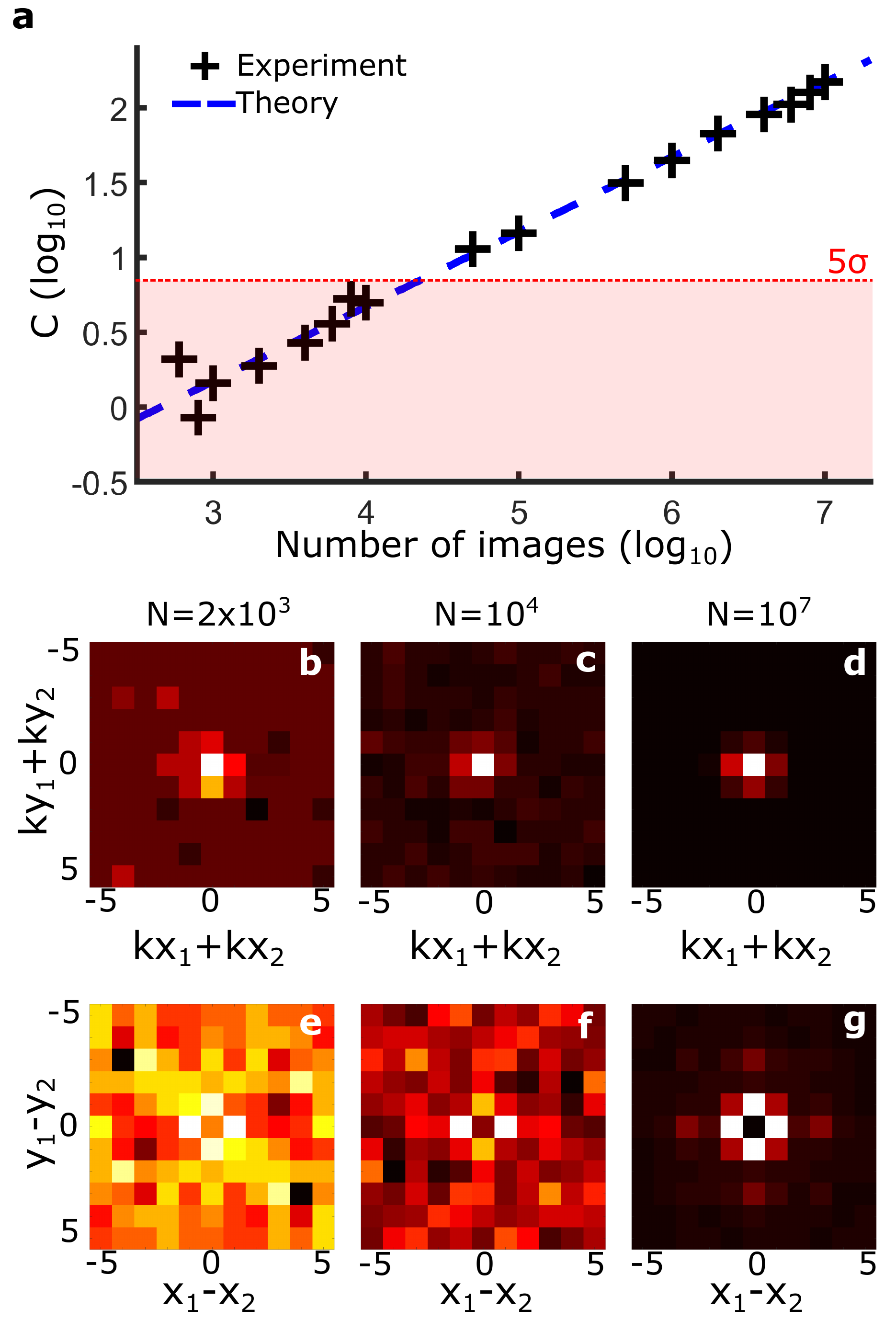}
		\caption{\label{Figure4} \textbf{Confidence level analysis}. \textbf{a,} $C$ values measured for different total number of frames $N$ (black crosses) together with a theoretical model of the form $0.047 \sqrt{N}$ (blue dashed line). \textbf{b-d}, Sum-coordinate projections of the JPD measured in the FF-configuration using \textbf{(b)} $2 \times 10^3$ frames, \textbf{(c)} $10^4$ frames and \textbf{(d)} $10^7$ frames. \textbf{e-g}, Minus-coordinate projection of the JPD measured in the NF configuration using \textbf{(e)} $2 \times 10^3$ frames, \textbf{(f)} $10^4$ frames and \textbf{(g)} $10^7$ frames.}
	\end{figure}
	\noindent\textbf{Confidence level analysis.} As detailed in the Methods section, $\sigma$ is calculated from the standard deviation of the noise surrounding the coincidence peaks in the sum- and minus-coordinates projections of the JPD. For a fixed exposure time and a constant source intensity, $\sigma$ depends only on the number of frames acquired to compute the JPD~\cite{reichert_optimizing_2018-3}. Figure~\ref{Figure4}a shows that the measured values of $C$ for different total number of frames $N$ (black crosses) are found to scale as $\sqrt{N}$ (blue dashed curve, see Methods). In particular, confident EPR violation ($C>5$) is not achieved for $N<2\times10^4$. Figures~\ref{Figure4}b-g show examples of sum- and minus-coordinate projections obtained for a total number of frames $N=2\times10^3$ (Figs \ref{Figure4}b and \ref{Figure4}e), $N=10^4$ (Figs. \ref{Figure4}c and \ref{Figure4}f) and $N=10^7$ (Figs. \ref{Figure4}d and \ref{Figure4}g). \BN{We clearly observe a decrease of the noise in the projection images with increasing number of frames. Indeed, the average over a larger $N$ in Eq.~\eqref{eq: JPD intensity} enables to obtain a better estimation of the accidental term, such that its subtraction yields a more precise JPD. Consequently, after subtraction of the accidentals, the signal-to-noise ratio of the JPD increases with the total number of frames, as does our confidence in the EPR violation. However, we observed that the position-correlation signal appears noisier than its momentum counterpart. This is because we do not detect a significant part of our signal originating from two photons incident on the same pixel. 
	Given that our SPAD camera cannot resolve photon number, we measure the weaker correlation signal from photons incident on adjacent pixels, owing to the relatively high 80\% fill-factor.}\\

	\begin{figure}[t]
	\centering\includegraphics[width=1 \columnwidth]{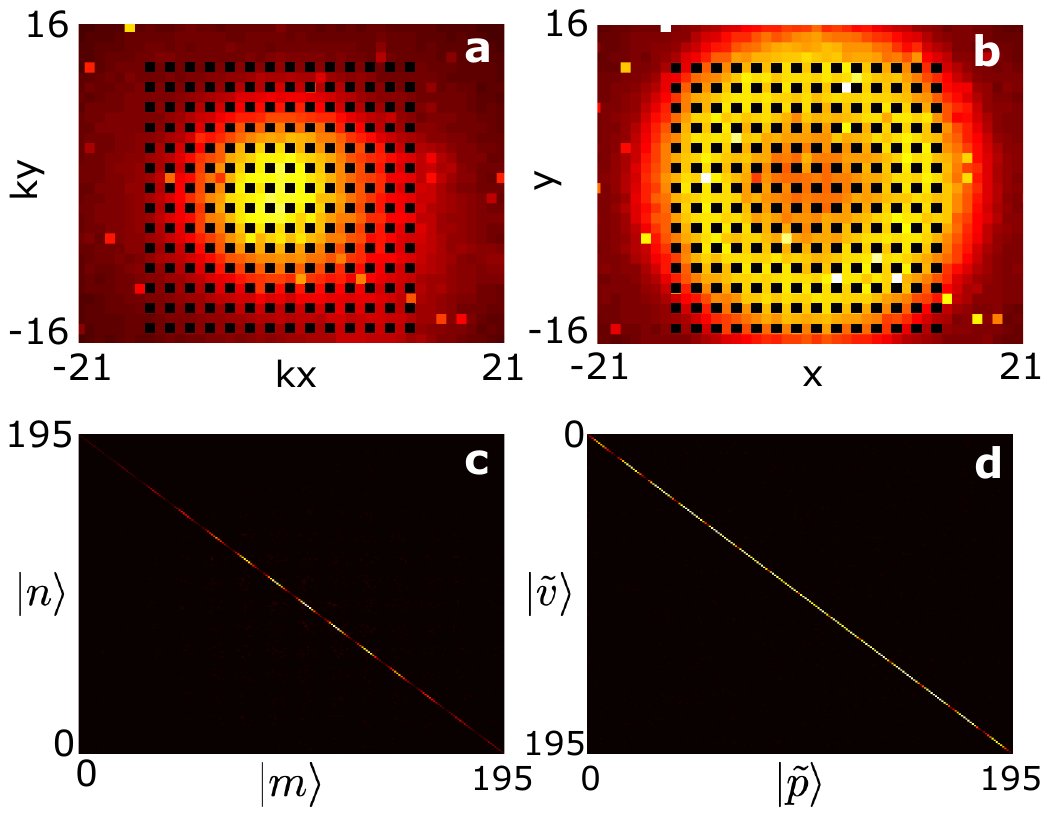}
	\caption{\label{Figure5} \BN{\textbf{Entanglement certification measurements}. In \textbf{a} near-field  and \textbf{b} far-field configurations, we select the same square grid of 14$\times$14 = 196 pixels, with a spacing of 1 pixel (grid of black squares). \textbf{c} and \textbf{d} show the measured correlations between all the pixels in the grid, with each pixel labelling the spatial coordinates of photon: $\ket{m}$ in the near-field (position) and $\ket{\tilde{p}}$ in the far-field (momentum). We calculated a lower bound $\tilde{F} = 0.252(9)$ of the fidelity with respect to a $d$~=~196 maximally entangled state, leading to a certified entanglement dimensionality of 48. Spatial coordinates are in pixels and the analysis was performed on a total of $10^7$ images.}}
\end{figure}

\BN{\noindent\textbf{Certification of high-dimensional entanglement.}
Measurements in two mutually unbiased bases (MUB) allow the use of a recently developed entanglement witness for certifying high-dimensional entanglement~\cite{Bavaresco2018}. In this respect, discrete position and momentum basis can be used as two MUBs and are accessible in our experimental setup using the NF- and FF-configurations~\cite{Erker2017}. As shown in Figures~\ref{Figure5}a and b, we then select $d=196$ pixels uniformly distributed over a central region in a grid of $14 \times 14$ pixels in both configurations. Modes associated with the chosen pixels are denoted $\{ \ket{m}\}_{m \in [\![0;d-1]\!]}$ (discrete position basis) and $\{ \ket{\tilde{p}}_{p \in [\![0;d-1]\!]} \}$ (discrete momentum basis). To certify entanglement dimensionality of the generated state $\rho$, correlation measurements are performed in the two MUBs to compute a lower bound for the fidelity of the state $\rho$ with respect to a maximally entangled target state $\ket{\Phi} = \frac{1}{\sqrt{d}} \sum_{m=0}^{d-1} \ket{mm}$. From the correlation matrices measured in Figs.~\ref{Figure5}.c and d, we obtained a lower bound value of the fidelity: $\tilde{F}(\rho,\Phi) =0.252(9)$ (see Methods). The entanglement dimensionality $d_{ent}$ that is certifiable with this method is the maximal $r$ such that  
\begin{equation}
\label{fidelitybound}
\tilde{F}(\rho,\Phi) > \frac{r-1}{d},
\end{equation}
allowing us to certify $d_{ent} = 48$ dimensions. Finally, it is essential to note that the total acquisition time for performing all the measurements (i.e. in the two MUBs) and certifying the entanglement dimension was only $140$ seconds. As a comparison, one day was required to certify a similar amount of dimensions of entanglement ($55$) using a state-of-the-art single-outcome projective measurement technique~\cite{Valencia2020}, representing a speed improvement of approximately $500 \times$ with our method.}\

	\section{Discussion}
	We have used a SPAD camera to characterise \BN{and quantify high-dimensional entanglement between photon pairs}. By measuring position and momentum correlations using $10^7$ intensity images, we showed a violation of an EPR criterion by 227 sigmas for an acquisition time of 140 s. While EPR violation has been demonstrated by acquiring very few frames with a highly-sensitive EMCCD camera~\cite{lantz_einstein-podolsky-rosen_2015}, quantum imaging approaches based on correlation measurements between spatially entangled photon pairs require the measurement of a large number of frames~\cite{reichert_massively_2018, Toninelli2019,Gregory2019, Defienne2019, defienne_adaptive_2018-1, defienne_entanglement-enabled_2019}, typically on the order of $10^6-10^7$. This is to ensure high enough SNR on the conditional projections to reconstruct the image by exploiting photon-pair correlations. Using an EMCCD for example, the reconstruction of the spatial JPD is performed at a frame rate of the order of $100$ frames/s, amounting to a total acquisition time of  many hours. Our ability to reduce  the JPD measurement time  by a factor of 1000$\times$ will allow quantum imaging proof-of-principle experiments to evolve towards practical applications. 
	
	\BN{Furthermore, we have certified high-dimensional entanglement up to $48$ dimensions in just 140 seconds, a significant development in quantum information processing with high-dimensional quantum states. While certifying high-dimensional entanglement is now routinely performed using single-outcome projective measurement techniques~\cite{Bavaresco2018, Schneeloch2019, Valencia2019, Valencia2020}, these approaches are tedious and prohibitively time-consuming. Our SPAD-camera-based technique outperforms these approaches in term of speed and scalability, with a demonstrated speed-up by more than a factor $500 \times$ and the potential access to up to $4\times10^6$  two-photon modes ($(32\times64)^2$) in parallel, indicating a key role to be played in future quantum information processing systems based on high-dimensional entanglement. Finally, a natural combination with recently developed spatial mode sorting devices~\cite{Fontaine2019,Brandt2020} holds promise for exploring high-dimensional entanglement in other types of spatial mode bases, including those carrying orbital angular momentum.}
	
	\section{Methods} 
	\noindent \textbf{Details on $\Gamma$ reconstruction.}
	Equation~\ref{eq: JPD intensity} enables the reconstruction of the spatial JPD from a finite number of frames $N$ acquired with the SPAD camera. This equation is derived from a theoretical model of photon pair detection detailed in~\cite{defienne_general_2018-2}. In this work, a link is established between the JPD and the measured frames at the limit $N \rightarrow + \infty$:
	\begin{equation}
	\label{equtotal}
	\Gamma(\vec{r_i},\vec{r_j}) = A \ln \left(1+\frac{\langle I(\vec{r_i}) I(\vec{r_j}) \rangle-\langle I(\vec{r_i}) \rangle \langle I(\vec{r_j}) \rangle}{(1-\langle I(\vec{r_i}) \rangle)(1-\langle I(\vec{r_i}) \rangle)}\right),
	\end{equation}
	where $A$ is a constant coefficient that depends on both the quantum efficiency of the sensor and the power of the pump laser, and
	\begin{eqnarray}
	\langle I(\vec{r_i}) I(\vec{r_j}) \rangle &=& \lim_{N \rightarrow + \infty} \frac{1}{N} \sum_{l=1}^{N} I_l(\vec{r_i}) I_l(\vec{r_j}), \\
	\langle I(\vec{r_i}) \rangle &=& \lim_{N \rightarrow + \infty} \frac{1}{N} \sum_{l=1}^{N} I_l(\vec{r_i}).
	\end{eqnarray}
	Equation~\ref{equtotal} is obtained under hypotheses~\cite{defienne_general_2018-2} that are all verified in our work, including that (i) the quantum efficiency is the same for all pixels of the sensor and (ii) the number of pairs produced by SPDC during the time an exposure follows a Poisson distribution~\cite{larchuk_statistics_1995}. Moreover, in our experiment the probability of detecting a photon per pixel per frame is much lower than one ($\langle I(\vec{r}) \rangle \ll 1$), which allows us to express Eq.~\ref{equtotal} as follows:
	\begin{equation}
	\label{equ3}
	\Gamma(\vec{r_i},\vec{r_j}) \approx \langle I(\vec{r_i}) I(\vec{r_j}) \rangle-\langle I(\vec{r_i}) \rangle \langle I(\vec{r_j}) \rangle.
	\end{equation}
	In the practical case where only a finite number of frames $N$ is measured, the first term on the right-hand side in Eq.~\ref{equ3} is estimated by multiplying pixel values within the same frame:
	\begin{equation}
	\label{equ4}
	\langle I(\vec{r_i}) I(\vec{r_j}) \rangle \approx \frac{1}{N}\sum_{l=1}^N I_l(\vec{r_i}) I_l(\vec{r_j}).
	\end{equation}
	The second term on the right-hand side in Eq.~\ref{equ3} is estimated by multiplying the averaged intensity values: 
	\begin{equation}
	\label{equ5}
	\langle I(\vec{r_i}) \rangle \langle I(\vec{r_j})\rangle \approx \frac{1}{N^2}\sum_{m,n=1}^{N} I_m(\vec{r_i}) I_{n}(\vec{r_j}).
	\end{equation}
	Combining Eqs.~\ref{equ3},~\ref{equ4} and \ref{equ5} finally leads to Eq~\ref{eq: JPD intensity}. \\\\
	
	\noindent \textbf{$\Delta\vec{r}$ and $\Delta\vec{k}$ measurements and uncertainties.}
	Transverse position and momentum  correlation widths, $\Delta\vec{r}$ and $\Delta\vec{k}$, are estimated by fitting the sum- and minus-coordinate projections of the JPD measured in FF- and NF-configurations by a Gaussian model~\cite{moreau_realization_2012,Edgar2012} of the form $f(r)=$ $a\exp(-r^2/2\Delta^2)$, where $a$ is a fitting parameter and $\Delta$ is the desired correlation width value ($\Delta\vec{r}$ or $\Delta\vec{k}$). Note that, in the case of the position correlation width measurement, the central point of the minus-coordinate projection is excluded from the fitted data. The presence of noise in the sum- and minus-coordinate images induce uncertainties on values $\Delta\vec{r}$ and $\Delta\vec{k}$ returned by the fitting process. The standard deviation of the noise $\Sigma$ is measured in an area composed of $40 \times 40$ pixels surrounding the central peak of coincidence. \BN{The link between the correlation width uncertainty $\delta_\Delta$ ($\delta_\Delta = \delta_{\Delta\vec{r}}$ or $\delta_{\Delta\vec{k}}$) and $\Sigma$ is given by calculating the value of $grad[f]$ a the position $r = \Delta$:
	\begin{equation}
	\left| \frac{df}{dr}(r=\Delta) \right| = \frac{a}{\Delta \sqrt{e}}.
	\label{derivative}
	\end{equation}
	and expanding it at the first order in $r$:
	\begin{equation}
	\delta f = \frac{a}{\Delta \sqrt{e}} \delta r.
	\label{deltaDelta}
	\end{equation}
	In our case, the variations $\delta f$ and $\delta r$ identify to the uncertainty quantities $\Sigma$ and $\delta_\Delta$, respectively, which finally leads to:
	\begin{equation}
	\delta_\Delta = \frac{\Sigma \sqrt{e}\Delta}{a}.
	\label{deltaDelta}
	\end{equation}
	}
	All correlation width values and uncertainties are expressed in the coordinate system of the crystal, after taking into consideration the magnifications introduced by the imaging systems detailed in Fig.~\ref{Figure1}.\\\\
	
	\noindent \textbf{Variation of the confidence level $C$  with the number of images $N$.}
	As defined in Eq.~\ref{equconfidence}, the confidence levels $C$ depends on both $\Delta\vec{r}\cdot\Delta\vec{k}$ and its uncertainty $\sigma$. For a given non-linear crystal and a stationary pump, $\Delta\vec{r}\cdot\Delta\vec{k}$ is constant while $\sigma$ depends on the quality of our measurement, including the total number of acquired frames $N$. To establish the theoretical link between $C$ and $N$, we first relate $\sigma$ to the uncertainties in position and momentum correlation widths ($\delta_{\Delta\vec{r}}$ and $\delta_{\Delta\vec{k}}$) by error propagation:
	\begin{equation}
	\sigma= \Delta\vec{r}\cdot\Delta\vec{k} \sqrt{\left(\frac{\delta_{\Delta\vec{r}}}{\Delta\vec{r}} \right)^2+\left(\frac{\delta_{\Delta\vec{k}}}{\Delta\vec{k}}\right)^2}.
	\end{equation}
	Then, we replace $\delta_{\Delta\vec{k}}$ and $\delta_{\Delta\vec{r}}$ using Eq.~\ref{deltaDelta} to show that $\sigma \propto \Sigma$. Finally, we use the fact that Reichert \textit{et al.}. have shown that $\Sigma \propto 1/ \sqrt{N}$ for a constant average pump power and a fixed exposure time~\cite{reichert_optimizing_2018-3}. Thus, we conclude that $C \propto \sqrt{N}.$ As shown in Fig.~\ref{Figure4}, this theoretical model fits  successfully with the experimental data  $(R^2 =0.998)$.\\\\
	
	\noindent \BN{\textbf{Effect of pixelation on the Gaussian fit.} As shown in Figures~\ref{Figure2} and~\ref{Figure3}, the correlation widths of the photon pairs produced in our experiment are on the order of the pixel size. This pixelation effect introduces uncertainties on the Gaussian fits and the extracted values of the correlation widths. While this error is difficult to quantify, it is essential to note it can only lead to an overestimation of the correlation width values. In the limiting case where the correlation width is much smaller than the size of a pixel (i.e. the correlation image has only one non-zero pixel surrounded by zeros), the Gaussian fit will always return a value that is approximately equal to $20\%$ of the pixel size. Therefore, the confidence value calculated from the correlation width values is a lower bound of the more accurate value that would be measured using a SPAD with more pixels. As a result, the pixelation effect cannot change our conclusion on the presence of entanglement. Simulations of the pixelation effect are provided in the supplementary document.\\\\
	
	\noindent \textbf{Discrete position and momentum basis.}
	In our experiment, we use discrete photonic transverse position and momentum bases given by a set pixels defined on the SPAD camera, which we refer to as the discrete position basis $\{ \ket{m}\}_{m \in [\![1;d]\!]}$ and discrete momentum basis $ \{ \ket{\tilde{p}} \} _{p \in [\![1;d]\!]} $. Our approach is based on the protocol proposed by Erker \textit{et al.}~\cite{Erker2017} where these bases are used as two mutually unbiased bases (MUBs) to certify high-dimensional entanglement. More precisely, they are linked according to:
	\begin{equation}
	\ket{\tilde{p}} = \frac{1}{\sqrt{d}} \sum_{m=0}^{d-1} \omega^{km} \ket{m} 
	\end{equation} 
	where $\omega = e^{2 \pi i / d}$. Experimentally, these bases are accessed using simple optics i.e. lenses to image or Fourier-image the output of the non-linear crystal producing the photon pairs. A subset of pixels is then selected in the illuminated areas of the sensor to optimise coincidence signals measured. In our case, we selected 196 pixels evenly separated from each other by $1$ pixel and located on a square grid of $14 \times 14$ pixels at the center of the sensor. The only difference between the scheme proposed by Eker \textit{et al.}. and our work is that only one image is produced on the camera in our case, against two in their proposal using a collinear beam splitter. This setup prevents us from accessing the coincidence rate at the same pixel (e.g. same spatial modes) as already pointed out in Figure~\ref{Figure3}. Rather, the coincidence rate of photon pairs in the same pixel is inferred by measuring the coincidence rate between each pixel and its neighbour. This inference leads to a lower value of the intra-pixel coincidence rate in the NF-configuration. This approximation therefore underestimates the actual entanglement witness value.\\\\

	\noindent \textbf{Derivation of the dimensionality witness.} 
To certify the presence of high-dimensional entanglement in the measured state $\rho$, we employ a recently developed witness that uses correlations in at two MUBs~\cite{Bavaresco2018}. In our experiment, the two MUBs are the discrete position basis $\{ \ket{m}\}_{m \in [\![1;d]\!]}$ and discrete momentum basis $ \{ \ket{\tilde{p}} \} _{p \in [\![1;d]\!]}$ (see previous methods section). Using only coincidence measurements in these two bases, one can determine a lower bound for the fidelity $F(\rho,\Phi)$ of the state $\rho$ to a pure bipartite maximally entangled target state $\ket{\Phi}$. Since the fidelity to a target entangled state also provides information about the dimensionality of entanglement, we use this bound for certifying the dimension of entanglement of the state produced in our experiment.

We consider a maximally entangled target state written as:
\begin{equation}
\ket{\Phi} = \frac{1}{\sqrt{d}}\sum_{m=0}^{d-1} \ket{mm}
\end{equation}
with $d=196$. The fidelity $F(\rho,\Phi)$ of the state $\rho$ to the target state $\ket{\Phi}$ is defined as:
\begin{eqnarray}
\label{fidelity}
F(\rho,\Phi) &&= \mbox{Tr} \left(\ket{\Phi} \bra{\Phi} \rho \right) \nonumber \\
&&= \sum_{m,n=0}^{d-1} \brakket{mm}{\rho}{nn} \nonumber \\
&& = F_1(\rho,\Phi)+F_2(\rho,\Phi)
\end{eqnarray}

where 
\begin{eqnarray}
 F_1(\rho,\Phi) &=& \sum_{m=0}^{d-1} \brakket{mm}{\rho}{mm} \label{F1} \\
 F_2(\rho,\Phi) &=& \sum_{m\neq n}^{d-1} \brakket{mm}{\rho}{nn} \label{F2}
\end{eqnarray}

The entanglement dimensionality can be deduced from the fidelity taking into account that for any state $\rho$ of Schmidt number $r \leq d$, the fidelity of Eq.~\eqref{fidelity} is bound by:
\begin{equation}
\label{fidelity}
F(\rho,\Phi) \leq B_r(\Phi) = \frac{r}{d} 
\end{equation}
Hence, any state with $F(\rho,\Phi) > B_r(\Phi)$ must have an entanglement dimensionality of at least $r+1$. Our goal is therefore to obtain a lower bound on the fidelity as large as possible for the target state whose Schmidt rank is as close as possible to the local dimension $d$. To achieve this experimentally, the method described in~\cite{Bavaresco2018} works the following way:\\
\noindent Step 1: Matrix elements $ \{ \brakket{mn}{\rho}{mn} \}_{m,n}$ are calculated from the coincidence counts $\{N_{mn}\}_{mn}$ measured in the discrete position basis via:
\begin{equation}
\brakket{mn}{\rho}{mn} = \frac{N_{mn}}{\sum_{k,l}N_{kl}} 
\end{equation}
These elements are shown in the matrix in Figure~\ref{Figure5}.d. They enable to calculate directly the term $F_1(\rho,\Phi) = 0.00170 (1)$ from the definition, Eq~\eqref{F1}.\\
\noindent Step 2: Matrix elements $ \{ \brakket{\tilde{p} \tilde{v}}{\rho}{\tilde{p} \tilde{v}} \}_{p,v}$ are calculated from the coincidence counts $\{\tilde{N}_{pv}\}_{pv}$ measured in the discrete momentum basis via:
\begin{equation}
\brakket{\tilde{p} \tilde{v}}{\rho}{\tilde{p} \tilde{v}} = \frac{\tilde{N}_{pv}}{\sum_{k,l}\tilde{N}_{kl}} 
\end{equation}
These elements are shown in the matrix in Figure~\ref{Figure5}.c. These matrix elements, together those of the discrete position basis, allow us to bound the fidelity term $F_2(\rho,\Phi)$. This lower bound $\tilde{F}_2(\rho,\Phi) = 0.250(9)$ is calculated via:
\begin{eqnarray}
&&\tilde{F}_2(\rho,\Phi) =  \sum_{p=0}^{d-1} \brakket{\tilde{p} \tilde{p}}{\rho}{\tilde{p} \tilde{p}} - \frac{1}{d} - \nonumber \\
&& \sum_{\substack{m \neq n',m \neq n  \\ n \neq n',n' \neq m' }} \gamma_{mnm'n'} \sqrt{\brakket{mn}{\rho}{mn} \brakket{m'n'}{\rho}{m'n'} } \label{f2tilde}
\end{eqnarray}
where the prefactor $\gamma_{mnm'n'}$ is given by
\begin{equation}
\gamma_{mnm'n'} = \left\{
    \begin{array}{ll}
        0 & \mbox{if } (m-m'-n+n') \mbox{ mod } d\neq0 \\
        \frac{1}{d} & \mbox{otherwise.}
    \end{array}
\right.
\end{equation}
A derivation of Eq.~\eqref{f2tilde} can be found in the Methods section of~\cite{Bavaresco2018}.\\
\noindent Step 3: A lower bound on entanglement is calculated as $\tilde{F}(\rho,\Phi) = F_1(\rho,\Phi)+\tilde{F}_2(\rho,\Phi) =0.252(9) \leq F(\rho,\Phi)$. This lower bound value is finally compared to the certification bound $B_r(\Phi)$ as
\begin{equation}
B_r(\Phi) < \tilde{F}(\rho,\Phi) \leq B_{r+1}(\Phi)
\end{equation}
thus certifying entanglement in $r+1$ dimensions. In our experiment we found, accounting for errors, $r=47$ and hence certified entanglement in $d_{ent}=48$ dimensions.}\\

\BN{\noindent{\textit{Assumptions:}} Using this approach, we note that no assumptions are directly made about the underlying quantum state $\rho$. However, an assumption is made about our measurement process. Indeed, by using Eq.~\eqref{eq: JPD intensity} to measure the JPD of photon pairs, we effectively perform a subtraction of accidental counts. Correcting for accidental coincidence is acceptable in our experiment since we trust our measurement devices and the final goal is only to assess the presence of entanglement and its dimension. However, such an assumption would not be acceptable in an adversarial scenario such as quantum key distribution as it is likely to compromise the security of the protocol. Furthermore, this assumption cannot be used if one wants to perform a loophole-free non-locality tests, because it is likely to violate the fair-sampling assumption.} \\
\section{Data availability} 
	The experimental data and codes that support the findings presented here are available from the corresponding authors upon reasonable request.


\begin{thebibliography}{60}%
		\makeatletter
		\providecommand \@ifxundefined [1]{%
			\@ifx{#1\undefined}
		}%
		\providecommand \@ifnum [1]{%
			\ifnum #1\expandafter \@firstoftwo
			\else \expandafter \@secondoftwo
			\fi
		}%
		\providecommand \@ifx [1]{%
			\ifx #1\expandafter \@firstoftwo
			\else \expandafter \@secondoftwo
			\fi
		}%
		\providecommand \natexlab [1]{#1}%
		\providecommand \enquote  [1]{``#1''}%
		\providecommand \bibnamefont  [1]{#1}%
		\providecommand \bibfnamefont [1]{#1}%
		\providecommand \citenamefont [1]{#1}%
		\providecommand \href@noop [0]{\@secondoftwo}%
		\providecommand \href [0]{\begingroup \@sanitize@url \@href}%
		\providecommand \@href[1]{\@@startlink{#1}\@@href}%
		\providecommand \@@href[1]{\endgroup#1\@@endlink}%
		\providecommand \@sanitize@url [0]{\catcode `\\12\catcode `\$12\catcode
			`\&12\catcode `\#12\catcode `\^12\catcode `\_12\catcode `\%12\relax}%
		\providecommand \@@startlink[1]{}%
		\providecommand \@@endlink[0]{}%
		\providecommand \url  [0]{\begingroup\@sanitize@url \@url }%
		\providecommand \@url [1]{\endgroup\@href {#1}{\urlprefix }}%
		\providecommand \urlprefix  [0]{URL }%
		\providecommand \Eprint [0]{\href }%
		\providecommand \doibase [0]{http://dx.doi.org/}%
		\providecommand \selectlanguage [0]{\@gobble}%
		\providecommand \bibinfo  [0]{\@secondoftwo}%
		\providecommand \bibfield  [0]{\@secondoftwo}%
		\providecommand \translation [1]{[#1]}%
		\providecommand \BibitemOpen [0]{}%
		\providecommand \bibitemStop [0]{}%
		\providecommand \bibitemNoStop [0]{.\EOS\space}%
		\providecommand \EOS [0]{\spacefactor3000\relax}%
		\providecommand \BibitemShut  [1]{\csname bibitem#1\endcsname}%
		\let\auto@bib@innerbib\@empty
		\bibitem [{\citenamefont {Migdall}\ \emph {et~al.}(2013)\citenamefont
			{Migdall}, \citenamefont {Polyakov}, \citenamefont {Fan},\ and\ \citenamefont
			{Bienfang}}]{Migdall_book}%
		\BibitemOpen
		\bibfield  {author} {\bibinfo {author} {\bibfnamefont {A.}~\bibnamefont
				{Migdall}}, \bibinfo {author} {\bibfnamefont {S.}~\bibnamefont {Polyakov}},
			\bibinfo {author} {\bibfnamefont {J.}~\bibnamefont {Fan}}, \ and\ \bibinfo
			{author} {\bibfnamefont {J.}~\bibnamefont {Bienfang}},\ } \bibinfo
		{title} {{Single-Photon Generation and Detection: Physics and Applications}},\ Experimental methods in the physical sciences\ \bibinfo
		{publisher} {Elsevier},\ \bibinfo {year} {2013}\BibitemShut {NoStop}%
		\bibitem [{\citenamefont {Hadfield}(2009)}]{Hadfield2009}%
		\BibitemOpen
		\bibfield  {author} {\bibinfo {author} {\bibfnamefont {R.~H.}\ \bibnamefont
				{Hadfield}},\ }\href {\doibase 10.1038/nphoton.2009.230} {\bibfield
			{journal} {\bibinfo  {journal} {Nature Photonics}\ }\textbf {\bibinfo
				{volume} {3}},\ \bibinfo {pages} {696} (\bibinfo {year} {2009})}\BibitemShut
		{NoStop}%
		\bibitem [{\citenamefont {Cova}\ \emph {et~al.}(1996)\citenamefont {Cova},
			\citenamefont {Ghioni}, \citenamefont {Lacaita}, \citenamefont {Samori},\
			and\ \citenamefont {Zappa}}]{Cova_avalanche_1996}%
		\BibitemOpen
		\bibfield  {author} {\bibinfo {author} {\bibfnamefont {S.}~\bibnamefont
				{Cova}}, \bibinfo {author} {\bibfnamefont {M.}~\bibnamefont {Ghioni}},
			\bibinfo {author} {\bibfnamefont {A.}~\bibnamefont {Lacaita}}, \bibinfo
			{author} {\bibfnamefont {C.}~\bibnamefont {Samori}}, \ and\ \bibinfo {author}
			{\bibfnamefont {F.}~\bibnamefont {Zappa}},\ }\href {\doibase
			10.1364/AO.35.001956} {\bibfield  {journal} {\bibinfo  {journal} {Applied
					Optics}\ }\textbf {\bibinfo {volume} {35}},\ \bibinfo {pages} {1956} (\bibinfo
			{year} {1996})}\BibitemShut {NoStop}%
		\bibitem [{\citenamefont {Rochas}\ \emph
			{et~al.}(2003{\natexlab{a}})\citenamefont {Rochas}, \citenamefont {Gani},
			\citenamefont {Furrer}, \citenamefont {Besse}, \citenamefont {Popovic},
			\citenamefont {Ribordy},\ and\ \citenamefont {Gisin}}]{Rochas2003}%
		\BibitemOpen
		\bibfield  {author} {\bibinfo {author} {\bibfnamefont {A.}~\bibnamefont
				{Rochas}}, \bibinfo {author} {\bibfnamefont {M.}~\bibnamefont {Gani}},
			\bibinfo {author} {\bibfnamefont {B.}~\bibnamefont {Furrer}}, \bibinfo
			{author} {\bibfnamefont {P.~A.}\ \bibnamefont {Besse}}, \bibinfo {author}
			{\bibfnamefont {R.~S.}\ \bibnamefont {Popovic}}, \bibinfo {author}
			{\bibfnamefont {G.}~\bibnamefont {Ribordy}}, \ and\ \bibinfo {author}
			{\bibfnamefont {N.}~\bibnamefont {Gisin}},\ }\href {\doibase
			10.1063/1.1584083} {\bibfield  {journal} {\bibinfo  {journal} {Review of
					Scientific Instruments}\ }\textbf {\bibinfo {volume} {74}},\ \bibinfo {pages}
			{3263} (\bibinfo {year} {2003}{\natexlab{a}})}\BibitemShut {NoStop}%
		\bibitem [{\citenamefont {Rochas}\ \emph
			{et~al.}(2003{\natexlab{b}})\citenamefont {Rochas}, \citenamefont {Gosch},
			\citenamefont {Serov}, \citenamefont {Besse}, \citenamefont {Popovic},
			\citenamefont {Lasser},\ and\ \citenamefont {Rigler}}]{Rochas2003b}%
		\BibitemOpen
		\bibfield  {author} {\bibinfo {author} {\bibfnamefont {A.}~\bibnamefont
				{Rochas}}, \bibinfo {author} {\bibfnamefont {M.}~\bibnamefont {Gosch}},
			\bibinfo {author} {\bibfnamefont {A.}~\bibnamefont {Serov}}, \bibinfo
			{author} {\bibfnamefont {P.}~\bibnamefont {Besse}}, \bibinfo {author}
			{\bibfnamefont {R.}~\bibnamefont {Popovic}}, \bibinfo {author} {\bibfnamefont
				{T.}~\bibnamefont {Lasser}}, \ and\ \bibinfo {author} {\bibfnamefont
				{R.}~\bibnamefont {Rigler}},\ }\href {\doibase 10.1109/LPT.2003.813387}
		{\bibfield  {journal} {\bibinfo  {journal} {IEEE Photonics Technology
					Letters}\ }\textbf {\bibinfo {volume} {15}},\ \bibinfo {pages} {963}
			(\bibinfo {year} {2003}{\natexlab{b}})}\BibitemShut {NoStop}%
		\bibitem [{\citenamefont {Bruschini}\ \emph {et~al.}(2019)\citenamefont
			{Bruschini}, \citenamefont {Homulle}, \citenamefont {Antolovic},
			\citenamefont {Burri},\ and\ \citenamefont {Charbon}}]{Bruschini2019}%
		\BibitemOpen
		\bibfield  {author} {\bibinfo {author} {\bibfnamefont {C.}~\bibnamefont
				{Bruschini}}, \bibinfo {author} {\bibfnamefont {H.}~\bibnamefont {Homulle}},
			\bibinfo {author} {\bibfnamefont {I.~M.}\ \bibnamefont {Antolovic}}, \bibinfo
			{author} {\bibfnamefont {S.}~\bibnamefont {Burri}}, \ and\ \bibinfo {author}
			{\bibfnamefont {E.}~\bibnamefont {Charbon}},\ }\href {\doibase
			10.1038/s41377-019-0191-5} {\bibfield  {journal} {\bibinfo  {journal} {Light:
					Science {\&} Applications}\ }\textbf {\bibinfo {volume} {8}},\ \bibinfo
			{pages} {87} (\bibinfo {year} {2019})}\ \BibitemShut {NoStop}%
		\bibitem [{\citenamefont {Li}\ \emph {et~al.}(2010)\citenamefont {Li},
			\citenamefont {Arlt}, \citenamefont {Richardson}, \citenamefont {Walker},
			\citenamefont {Buts}, \citenamefont {Stoppa}, \citenamefont {Charbon},\ and\
			\citenamefont {Henderson}}]{Li2010}%
		\BibitemOpen
		\bibfield  {author} {\bibinfo {author} {\bibfnamefont {D.-U.}\ \bibnamefont
				{Li}}, \bibinfo {author} {\bibfnamefont {J.}~\bibnamefont {Arlt}}, \bibinfo
			{author} {\bibfnamefont {J.}~\bibnamefont {Richardson}}, \bibinfo {author}
			{\bibfnamefont {R.}~\bibnamefont {Walker}}, \bibinfo {author} {\bibfnamefont
				{A.}~\bibnamefont {Buts}}, \bibinfo {author} {\bibfnamefont {D.}~\bibnamefont
				{Stoppa}}, \bibinfo {author} {\bibfnamefont {E.}~\bibnamefont {Charbon}}, \
			and\ \bibinfo {author} {\bibfnamefont {R.}~\bibnamefont {Henderson}},\ }\href
		{\doibase 10.1364/OE.18.010257} {\bibfield  {journal} {\bibinfo  {journal}
				{Optics Express}\ }\textbf {\bibinfo {volume} {18}},\ \bibinfo {pages}
			{10257} (\bibinfo {year} {2010})}\BibitemShut {NoStop}%
		\bibitem [{\citenamefont {Henderson}\ \emph {et~al.}(2018)\citenamefont
			{Henderson}, \citenamefont {Johnston}, \citenamefont {Chen}, \citenamefont
			{Li}, \citenamefont {Hungerford}, \citenamefont {Hirsch}, \citenamefont
			{McLoskey}, \citenamefont {Yip},\ and\ \citenamefont
			{Birch}}]{Henderson2018}%
		\BibitemOpen
		\bibfield  {author} {\bibinfo {author} {\bibfnamefont {R.~K.}\ \bibnamefont
				{Henderson}}, \bibinfo {author} {\bibfnamefont {N.}~\bibnamefont {Johnston}},
			\bibinfo {author} {\bibfnamefont {H.}~\bibnamefont {Chen}}, \bibinfo {author}
			{\bibfnamefont {D.~D.~U.}\ \bibnamefont {Li}}, \bibinfo {author}
			{\bibfnamefont {G.}~\bibnamefont {Hungerford}}, \bibinfo {author}
			{\bibfnamefont {R.}~\bibnamefont {Hirsch}}, \bibinfo {author} {\bibfnamefont
				{D.}~\bibnamefont {McLoskey}}, \bibinfo {author} {\bibfnamefont
				{P.}~\bibnamefont {Yip}}, \ and\ \bibinfo {author} {\bibfnamefont {D.~J.}\
				\bibnamefont {Birch}},\ }\href {\doibase 10.1109/ESSCIRC.2018.8494330}
		{\bibfield  {journal} {\bibinfo  {journal} {ESSCIRC 2018 - IEEE 44th European
					Solid State Circuits Conference},\ \bibinfo {pages} {54}} (\bibinfo {year}
			{2018})}\BibitemShut {NoStop}%
		\bibitem [{\citenamefont {Bronzi}\ \emph {et~al.}(2014)\citenamefont {Bronzi},
			\citenamefont {Villa}, \citenamefont {Tisa}, \citenamefont {Tosi},
			\citenamefont {Zappa}, \citenamefont {Durini}, \citenamefont {Weyers},\ and\
			\citenamefont {Brockherde}}]{Bronzi2014}%
		\BibitemOpen
		\bibfield  {author} {\bibinfo {author} {\bibfnamefont {D.}~\bibnamefont
				{Bronzi}}, \bibinfo {author} {\bibfnamefont {F.}~\bibnamefont {Villa}},
			\bibinfo {author} {\bibfnamefont {S.}~\bibnamefont {Tisa}}, \bibinfo {author}
			{\bibfnamefont {A.}~\bibnamefont {Tosi}}, \bibinfo {author} {\bibfnamefont
				{F.}~\bibnamefont {Zappa}}, \bibinfo {author} {\bibfnamefont
				{D.}~\bibnamefont {Durini}}, \bibinfo {author} {\bibfnamefont
				{S.}~\bibnamefont {Weyers}}, \ and\ \bibinfo {author} {\bibfnamefont
				{W.}~\bibnamefont {Brockherde}},\ }\href {\doibase
			10.1109/JSTQE.2014.2341562} {\bibfield  {journal} {\bibinfo  {journal} {IEEE
					Journal of Selected Topics in Quantum Electronics}\ }\textbf {\bibinfo
				{volume} {20}},\ \bibinfo {pages} {354} (\bibinfo {year} {2014})}\BibitemShut
		{NoStop}%
		\bibitem [{\citenamefont {Finlayson}\ \emph {et~al.}(2018)\citenamefont
			{Finlayson}, \citenamefont {Gyongy}, \citenamefont {Johnston}, \citenamefont
			{Calder}, \citenamefont {{Al Abbas}}, \citenamefont {Erdogan}, \citenamefont
			{Henderson}, \citenamefont {Dutton},\ and\ \citenamefont
			{Walker}}]{Finlayson2018}%
		\BibitemOpen
		\bibfield  {author} {\bibinfo {author} {\bibfnamefont {I.}~\bibnamefont {Gyongy}},
			\bibinfo {author}	{\bibfnamefont {T.}~\bibnamefont {{Al Abbas}}},
			\bibinfo {author} {\bibfnamefont {N.}~\bibnamefont
				{Finlayson}}, \bibinfo {author} {\bibfnamefont {N.}~\bibnamefont {Johnston}}, \bibinfo
			{author} {\bibfnamefont {N.}~\bibnamefont {Calder}}, \bibinfo {author}
			{\bibfnamefont {A.}~\bibnamefont {Erdogan}}, \bibinfo {author} {\bibfnamefont
				{N.~W.}~\bibnamefont {Dutton}},\ \bibinfo {author} {\bibfnamefont
				{R.}~\bibnamefont {Walker}}, \ and\ \bibinfo {author} {\bibfnamefont
				{R.~K.}~\bibnamefont {Henderson}},\ }\href {\doibase 10.1117/12.2501977} {\bibfield
			{journal} {\bibinfo  {journal} {Emerging Imaging and Sensing Technologies
					for Security and Defence III; and Unmanned Sensors, Systems, and
					Countermeasures}\ }\textbf {\bibinfo {volume} {10799}},\ \bibinfo {pages}
			{1079907} (\bibinfo {year} {2018})}\BibitemShut {NoStop}%
		\bibitem [{\citenamefont {Lindner}\ \emph {et~al.}(2018)\citenamefont
			{Lindner}, \citenamefont {Zhang}, \citenamefont {Antolovic}, \citenamefont
			{Wolf},\ and\ \citenamefont {Charbon}}]{Lindner2018}%
		\BibitemOpen
		\bibfield  {author} {\bibinfo {author} {\bibfnamefont {S.}~\bibnamefont
				{Lindner}}, \bibinfo {author} {\bibfnamefont {C.}~\bibnamefont {Zhang}},
			\bibinfo {author} {\bibfnamefont {I.~M.}\ \bibnamefont {Antolovic}}, \bibinfo
			{author} {\bibfnamefont {M.}~\bibnamefont {Wolf}}, \ and\ \bibinfo {author}
			{\bibfnamefont {E.}~\bibnamefont {Charbon}},\ }\ \href {\doibase
			10.1109/VLSIC.2018.8502386} { {\bibinfo {journal} {2018 IEEE Symposium
					on VLSI Circuits,}}} \bibinfo {pages}
		{69} (\bibinfo {year} {2018})\BibitemShut {NoStop}%
		\bibitem [{\citenamefont {Henderson}\ \emph {et~al.}(2019)\citenamefont
			{Henderson}, \citenamefont {Johnston}, \citenamefont {Hutchings},
			\citenamefont {Gyongy}, \citenamefont {Abbas}, \citenamefont {Dutton},
			\citenamefont {Tyler}, \citenamefont {Chan},\ and\ \citenamefont
			{Leach}}]{Henderson2019}%
		\BibitemOpen
		\bibfield  {author} {\bibinfo {author} {\bibfnamefont {R.~K.}\ \bibnamefont
				{Henderson}}, \bibinfo {author} {\bibfnamefont {N.}~\bibnamefont {Johnston}},
			\bibinfo {author} {\bibfnamefont {S.~W.}\ \bibnamefont {Hutchings}}, \bibinfo
			{author} {\bibfnamefont {I.}~\bibnamefont {Gyongy}}, \bibinfo {author}
			{\bibfnamefont {T.~A.}\ \bibnamefont {Abbas}}, \bibinfo {author}
			{\bibfnamefont {N.}~\bibnamefont {Dutton}}, \bibinfo {author} {\bibfnamefont
				{M.}~\bibnamefont {Tyler}}, \bibinfo {author} {\bibfnamefont
				{S.}~\bibnamefont {Chan}}, \ and\ \bibinfo {author} {\bibfnamefont
				{J.}~\bibnamefont {Leach}},\ }in \href {\doibase 10.1109/ISSCC.2019.8662355}
		{\bibfield  {journal} {\bibinfo  {journal} {IEEE
					International Solid-State Circuits Conference}},\ \bibinfo {pages} {106} (\bibinfo {year} {2019})}\BibitemShut
		{NoStop}%
		\bibitem [{\citenamefont {Gariepy}\ \emph {et~al.}(2016)\citenamefont
			{Gariepy}, \citenamefont {Tonolini}, \citenamefont {Henderson}, \citenamefont
			{Leach},\ and\ \citenamefont {Faccio}}]{Gariepy2016}%
		\BibitemOpen
		\bibfield  {author} {\bibinfo {author} {\bibfnamefont {G.}~\bibnamefont
				{Gariepy}}, \bibinfo {author} {\bibfnamefont {F.}~\bibnamefont {Tonolini}},
			\bibinfo {author} {\bibfnamefont {R.}~\bibnamefont {Henderson}}, \bibinfo
			{author} {\bibfnamefont {J.}~\bibnamefont {Leach}}, \ and\ \bibinfo {author}
			{\bibfnamefont {D.}~\bibnamefont {Faccio}},\ }\href {\doibase
			10.1038/nphoton.2015.234} {\bibfield  {journal} {\bibinfo  {journal} {Nature
					Photonics}\ }\textbf {\bibinfo {volume} {10}},\ \bibinfo {pages} {23}
			(\bibinfo {year} {2016})}\BibitemShut {NoStop}%
			\bibitem [{\citenamefont {Faccio}\ \emph {et~al.}(2016)\citenamefont
			{Faccio}, \citenamefont {Velten},\ and\ \citenamefont {Wetzstein}}]{NLOS2020}%
		\BibitemOpen
		\bibfield  {author} {\bibinfo {author} {\bibfnamefont {D.}~\bibnamefont
				{Faccio}}, \bibinfo {author} {\bibfnamefont {A.}~\bibnamefont {Velten}}, \ and\
			\bibinfo {author} {\bibfnamefont {G.}~\bibnamefont {Wetzstein}},\ }\href {\doibase
			10.1038/s42254-020-0174-8} {\bibfield  {journal} {\bibinfo  {journal} {Nat.
					Rev. Phys.}\ }\textbf {\bibinfo {volume} {2}},\ \bibinfo {pages} {318}
			(\bibinfo {year} {2020})}\BibitemShut {NoStop}%
		\bibitem [{\citenamefont {Lyons}\ \emph {et~al.}(2019)\citenamefont {Lyons},
			\citenamefont {Tonolini}, \citenamefont {Boccolini}, \citenamefont {Repetti},
			\citenamefont {Henderson}, \citenamefont {Wiaux},\ and\ \citenamefont
			{Faccio}}]{Lyons2019b}%
		\BibitemOpen
		\bibfield  {author} {\bibinfo {author} {\bibfnamefont {A.}~\bibnamefont
				{Lyons}}, \bibinfo {author} {\bibfnamefont {F.}~\bibnamefont {Tonolini}},
			\bibinfo {author} {\bibfnamefont {A.}~\bibnamefont {Boccolini}}, \bibinfo
			{author} {\bibfnamefont {A.}~\bibnamefont {Repetti}}, \bibinfo {author}
			{\bibfnamefont {R.}~\bibnamefont {Henderson}}, \bibinfo {author}
			{\bibfnamefont {Y.}~\bibnamefont {Wiaux}}, \ and\ \bibinfo {author}
			{\bibfnamefont {D.}~\bibnamefont {Faccio}},\ }\href {\doibase
			10.1038/s41566-019-0439-x} {\bibfield  {journal} {\bibinfo  {journal} {Nature
					Photonics}\ }\textbf {\bibinfo {volume} {13}},\ \bibinfo {pages} {575}
			(\bibinfo {year} {2019})}\BibitemShut {NoStop}%
			\bibitem [{\citenamefont {Lubin}\ \emph {et~al.}(2019)\citenamefont {Lubin},
			\citenamefont {Tenne}, \citenamefont {Antolovic}, \citenamefont {Charbon},
			\citenamefont {Bruschini} and\ \citenamefont
			{Oron}}]{Lubin2019}%
		\BibitemOpen
		\bibfield  {author} {\bibinfo {author} {\bibfnamefont {G.}~\bibnamefont
				{Lubin}}, \bibinfo {author} {\bibfnamefont {R.}~\bibnamefont {Tenne}},
			\bibinfo {author} {\bibfnamefont {I.M.}~\bibnamefont {Antolovic}}, \bibinfo
			{author} {\bibfnamefont {E.}~\bibnamefont {Charbon}}, \bibinfo {author}
			{\bibfnamefont {C.}~\bibnamefont {Bruschini}}, \ and\ \bibinfo {author}
			{\bibfnamefont {D.}~\bibnamefont {Oron}},\ }\href {\doibase
			10.1364/OE.27.032863} {\bibfield  {journal} {\bibinfo  {journal} {Optics Express}\ }\textbf {\bibinfo {volume} {27}}, {\bibinfo {number} {23}}, \ \bibinfo {pages} {32863--32882}
			(\bibinfo {year} {2019})}\BibitemShut {NoStop}%
		\bibitem [{\citenamefont {Unternährer}\ \emph {et~al.}(2016)\citenamefont {Unternährer},
			\citenamefont {Bessire}, \citenamefont {Gasparini}, \citenamefont {Stoppa},
			and\ \citenamefont{Stefanov}}]{Unternahrer2016}%
		\BibitemOpen
		\bibfield  {author} {\bibinfo {author} {\bibfnamefont {M.}~\bibnamefont
				{Unternährer}}, \bibinfo {author} {\bibfnamefont {B.}~\bibnamefont {Bessire}},
			\bibinfo {author} {\bibfnamefont {L.}~\bibnamefont {Gasparini}}, \bibinfo
			{author} {\bibfnamefont {D.}~\bibnamefont {Stoppa}} \ and\ \bibinfo {author}
			{\bibfnamefont {A.}~\bibnamefont {Stefanov}},\ }\href {\doibase
			10.1364/OE.24.028829} {\bibfield  {journal} {\bibinfo  {journal} {Optics Express}\ }\textbf {\bibinfo {volume} {24}}, {\bibinfo {number} {25}}, \ \bibinfo {pages} {32863}
			(\bibinfo {year} {2016})}\BibitemShut {NoStop}%
		\bibitem [{\citenamefont {Gariepy}\ \emph
			{et~al.}(2015{\natexlab{a}})\citenamefont {Gariepy}, \citenamefont
			{Krstaji{\'{c}}}, \citenamefont {Henderson}, \citenamefont {Li},
			\citenamefont {Thomson}, \citenamefont {Buller}, \citenamefont {Heshmat},
			\citenamefont {Raskar}, \citenamefont {Leach},\ and\ \citenamefont
			{Faccio}}]{Gariepy2015}%
		\BibitemOpen
		\bibfield  {author} {\bibinfo {author} {\bibfnamefont {G.}~\bibnamefont
				{Gariepy}}, \bibinfo {author} {\bibfnamefont {N.}~\bibnamefont
				{Krstaji{\'{c}}}}, \bibinfo {author} {\bibfnamefont {R.}~\bibnamefont
				{Henderson}}, \bibinfo {author} {\bibfnamefont {C.}~\bibnamefont {Li}},
			\bibinfo {author} {\bibfnamefont {R.~R.}\ \bibnamefont {Thomson}}, \bibinfo
			{author} {\bibfnamefont {G.~S.}\ \bibnamefont {Buller}}, \bibinfo {author}
			{\bibfnamefont {B.}~\bibnamefont {Heshmat}}, \bibinfo {author} {\bibfnamefont
				{R.}~\bibnamefont {Raskar}}, \bibinfo {author} {\bibfnamefont
				{J.}~\bibnamefont {Leach}}, \ and\ \bibinfo {author} {\bibfnamefont
				{D.}~\bibnamefont {Faccio}},\ }\href {\doibase 10.1038/ncomms7021} {\bibfield
			{journal} {\bibinfo  {journal} {Nature Communications}\ }\textbf {\bibinfo
				{volume} {6}},\ \bibinfo {pages} {6021} (\bibinfo {year}
			{2015}{\natexlab{a}})}\BibitemShut {NoStop}%
		\bibitem [{\citenamefont {Jost}\ \emph
			{et~al.}(1998{\natexlab{a}})\citenamefont {Jost}, \citenamefont
			{Sergienko}, \citenamefont
			{Abouraddy}, \citenamefont
			{Saleh},\ and\ \citenamefont {Teich}}]{Jost1998}%
		\BibitemOpen
		\bibfield  {author} {\bibinfo {author} {\bibfnamefont {B.~M.}~\bibnamefont
				{Jost}},  \bibinfo {author} {\bibfnamefont {A.~V.}~\bibnamefont {Sergienko}}, \bibinfo {author} {\bibfnamefont {A.~F.}~\bibnamefont {Abouraddy}}, \bibinfo {author} {\bibfnamefont {B.~E.~A.}~\bibnamefont {Saleh}},  \ and\ \bibinfo {author} {\bibfnamefont {M.~C.}\ \bibnamefont {Teich}},\
		}\href{\doibase 10.1364/OE.3.000081} {\bibfield  {journal} {\bibinfo  {journal} {Optics Express}}\ \textbf {\bibinfo {volume} {3}},\ \bibinfo {pages} {81} (\bibinfo {year} {1998})}
		\BibitemShut {NoStop}%
		\bibitem [{\citenamefont {Abouraddy}\ \emph
			{et~al.}(1998{\natexlab{a}})\citenamefont {Jost}, \citenamefont
			{Nasr}, \citenamefont
			{Saleh}, \citenamefont
			{Sergienko},\ and\ \citenamefont {Teich}}]{Abouraddy2001}%
		\BibitemOpen
		\bibfield  {author} {\bibinfo {author} {\bibfnamefont {A.~F.}~\bibnamefont
				{Abouraddy}},  \bibinfo {author} {\bibfnamefont {M.~B.}~\bibnamefont {Nasr}}, \bibinfo {author} {\bibfnamefont {B.~E.~A.}~\bibnamefont {Saleh}}, \bibinfo {author} {\bibfnamefont {A.~V.}~\bibnamefont {Sergienko}},  \ and\ \bibinfo {author} {\bibfnamefont {M.~C.}\ \bibnamefont {Teich}},\
		}\href{\doibase 110.1103/PhysRevA.63.063803} {\bibfield  {journal} {\bibinfo  {journal} {Physical Review A}}\ \textbf {\bibinfo {volume} {63}},\ \bibinfo {pages} {063803} (\bibinfo {year} {2001})}
		\BibitemShut {NoStop}%
		\bibitem [{\citenamefont {Pittman}\ \emph {et~al.}(1995)\citenamefont
			{Pittman}, \citenamefont {Shih}, \citenamefont {Strekalov},\ and\
			\citenamefont {Sergienko}}]{Pittman1995}%
		\BibitemOpen
		\bibfield  {author} {\bibinfo {author} {\bibfnamefont {T.~B.}\ \bibnamefont
				{Pittman}}, \bibinfo {author} {\bibfnamefont {Y.~H.}\ \bibnamefont {Shih}},
			\bibinfo {author} {\bibfnamefont {D.~V.}\ \bibnamefont {Strekalov}}, \ and\
			\bibinfo {author} {\bibfnamefont {A.~V.}\ \bibnamefont {Sergienko}},\ }\href
		{\doibase 10.1103/PhysRevA.52.R3429} {\bibfield  {journal} {\bibinfo
				{journal} {Physical Review A}\ }\textbf {\bibinfo {volume} {52}},\ \bibinfo
			{pages} {R3429} (\bibinfo {year} {1995})}\BibitemShut {NoStop}%
		\bibitem [{\citenamefont {Moreau}\ \emph {et~al.}(2019)\citenamefont {Moreau},
			\citenamefont {Toninelli}, \citenamefont {Gregory}, \citenamefont {Aspden},
			\citenamefont {Morris},\ and\ \citenamefont {Padgett}}]{Moreau2019}%
		\BibitemOpen
		\bibfield  {author} {\bibinfo {author} {\bibfnamefont {P.-A.}\ \bibnamefont
				{Moreau}}, \bibinfo {author} {\bibfnamefont {E.}~\bibnamefont {Toninelli}},
			\bibinfo {author} {\bibfnamefont {T.}~\bibnamefont {Gregory}}, \bibinfo
			{author} {\bibfnamefont {R.~S.}\ \bibnamefont {Aspden}}, \bibinfo {author}
			{\bibfnamefont {P.~A.}\ \bibnamefont {Morris}}, \ and\ \bibinfo {author}
			{\bibfnamefont {M.~J.}\ \bibnamefont {Padgett}},\ }\href {\doibase
			10.1126/sciadv.aaw2563} {\bibfield  {journal} {\bibinfo  {journal} {Science
					Advances}\ }\textbf {\bibinfo {volume} {5}},\ \bibinfo {pages} {eaaw2563}
			(\bibinfo {year} {2019})}\BibitemShut {NoStop}%
		\bibitem [{\citenamefont {Xu}\ \emph {et~al.}(2015)\citenamefont {Xu},
			\citenamefont {Song}, \citenamefont {Li}, \citenamefont {Zhang},
			\citenamefont {Wang}, \citenamefont {Xiong},\ and\ \citenamefont
			{Wang}}]{Xu2015}%
		\BibitemOpen
		\bibfield  {author} {\bibinfo {author} {\bibfnamefont {D.-Q.}\ \bibnamefont
				{Xu}}, \bibinfo {author} {\bibfnamefont {X.-B.}\ \bibnamefont {Song}},
			\bibinfo {author} {\bibfnamefont {H.-G.}\ \bibnamefont {Li}}, \bibinfo
			{author} {\bibfnamefont {D.-J.}\ \bibnamefont {Zhang}}, \bibinfo {author}
			{\bibfnamefont {H.-B.}\ \bibnamefont {Wang}}, \bibinfo {author}
			{\bibfnamefont {J.}~\bibnamefont {Xiong}}, \ and\ \bibinfo {author}
			{\bibfnamefont {K.}~\bibnamefont {Wang}},\ }\href {\doibase
			10.1063/1.4919131} {\bibfield  {journal} {\bibinfo  {journal} {Applied
					Physics Letters}\ }\textbf {\bibinfo {volume} {106}},\ \bibinfo {pages}
			{171104} (\bibinfo {year} {2015})}\BibitemShut {NoStop}%
		\bibitem [{\citenamefont {Toninelli}\ \emph {et~al.}(2019)\citenamefont
			{Toninelli}, \citenamefont {Moreau}, \citenamefont {Gregory}, \citenamefont
			{Mihalyi}, \citenamefont {Edgar}, \citenamefont {Radwell},\ and\
			\citenamefont {Padgett}}]{Toninelli2019}%
		\BibitemOpen
		\bibfield  {author} {\bibinfo {author} {\bibfnamefont {E.}~\bibnamefont
				{Toninelli}}, \bibinfo {author} {\bibfnamefont {P.-A.}\ \bibnamefont
				{Moreau}}, \bibinfo {author} {\bibfnamefont {T.}~\bibnamefont {Gregory}},
			\bibinfo {author} {\bibfnamefont {A.}~\bibnamefont {Mihalyi}}, \bibinfo
			{author} {\bibfnamefont {M.}~\bibnamefont {Edgar}}, \bibinfo {author}
			{\bibfnamefont {N.}~\bibnamefont {Radwell}}, \ and\ \bibinfo {author}
			{\bibfnamefont {M.}~\bibnamefont {Padgett}},\ }\href {\doibase
			10.1364/OPTICA.6.000347} {\bibfield  {journal} {\bibinfo  {journal} {Optica}\
			}\textbf {\bibinfo {volume} {6}},\ \bibinfo {pages} {347} (\bibinfo {year}
			{2019})}\BibitemShut {NoStop}%
		\bibitem [{\citenamefont {Zhang}\ \emph {et~al.}(2020)\citenamefont {Zhang},
			\citenamefont {England}, \citenamefont {Nomerotski}, \citenamefont {Svihra}, \citenamefont {Ferrante},  \citenamefont {Hockett},
			and\ \citenamefont{Sussman}}]{Zhang2020}%
		\BibitemOpen
		\bibfield  {author} {\bibinfo {author} {\bibfnamefont {Y.}~\bibnamefont
				{Zhang}}, \bibinfo {author} {\bibfnamefont {D.}~\bibnamefont {England}},
			\bibinfo {author} {\bibfnamefont {A.}~\bibnamefont {Nomerotski}}, \bibinfo
			{author} {\bibfnamefont {P.}~\bibnamefont {Svihra}}  \bibinfo
			{author} {\bibfnamefont {S.}~\bibnamefont {Ferrante}} \bibinfo
			{author} {\bibfnamefont {B.}~\bibnamefont {Hockett}}\ and\ \bibinfo {author}
			{\bibfnamefont {B.}~\bibnamefont {Sussman}},\ }\href {\doibase
			10.1103/PhysRevA.101.053808} {\bibfield  {journal} {\bibinfo  {journal} {Physical Review A}\ }\textbf {\bibinfo {volume} {101}}, \ \bibinfo {pages} {053808}
			(\bibinfo {year} {2020})}\BibitemShut {NoStop}%
		\bibitem [{\citenamefont {Defienne}\ \emph
			{et~al.}(2019{\natexlab{a}})\citenamefont {Defienne}, \citenamefont
			{Reichert}, \citenamefont {Fleischer},\ and\ \citenamefont
			{Faccio}}]{Defienne2019}%
		\BibitemOpen
		\bibfield  {author} {\bibinfo {author} {\bibfnamefont {H.}~\bibnamefont
				{Defienne}}, \bibinfo {author} {\bibfnamefont {M.}~\bibnamefont {Reichert}},
			\bibinfo {author} {\bibfnamefont {J.~W.}\ \bibnamefont {Fleischer}}, \ and\
			\bibinfo {author} {\bibfnamefont {D.}~\bibnamefont {Faccio}},\ }\href
		{\doibase 10.1126/sciadv.aax0307} {\bibfield  {journal} {\bibinfo  {journal}
				{Science Advances}\ }\textbf {\bibinfo {volume} {5}},\ \bibinfo {pages}
			{eaax0307} (\bibinfo {year} {2019}{\natexlab{a}})}\BibitemShut {NoStop}%
		\bibitem [{\citenamefont {Gregory}\ \emph {et~al.}(2019)\citenamefont
			{Gregory}, \citenamefont {Toninelli},\ and\ \citenamefont
			{Padgett}}]{Gregory2019}%
		\BibitemOpen
		\bibfield  {author} {\bibinfo {author} {\bibfnamefont {T.}~\bibnamefont
				{Gregory}},  \bibinfo {author} {\bibfnamefont {P-A.}~\bibnamefont {Moreau}}, \bibinfo {author} {\bibfnamefont {E.}~\bibnamefont {Toninelli}},
			\ and\ \bibinfo {author} {\bibfnamefont {M.~J.}\ \bibnamefont {Padgett}},\
		}\href {\doibase 10.1126/sciadv.aay2652} {\bibfield  {journal} {\bibinfo  {journal} {Science Advances}} \textbf
		{\bibinfo {volume} {6}},\ \bibinfo {pages} {eaay2652} \  (\bibinfo {year} {2020})}
		\BibitemShut {NoStop}%
		\bibitem [{\citenamefont {Ianzano}\ \emph {et~al.}(2020)\citenamefont
			{Ianzano}, \citenamefont {Svihra}, \citenamefont {Flament}, \citenamefont
			{Hardy}, \citenamefont {Cui}, \citenamefont {Nomerotski},\ and\ \citenamefont
			{Figueroa}}]{Ianzano2020}%
		\BibitemOpen
		\bibfield  {author} {\bibinfo {author} {\bibfnamefont {C.}~\bibnamefont
				{Ianzano}}, \bibinfo {author} {\bibfnamefont {P.}~\bibnamefont {Svihra}},
			\bibinfo {author} {\bibfnamefont {M.}~\bibnamefont {Flament}}, \bibinfo
			{author} {\bibfnamefont {A.}~\bibnamefont {Hardy}}, \bibinfo {author}
			{\bibfnamefont {G.}~\bibnamefont {Cui}}, \bibinfo {author} {\bibfnamefont
				{A.}~\bibnamefont {Nomerotski}}, \ and\ \bibinfo {author} {\bibfnamefont
				{E.}~\bibnamefont {Figueroa}},\ }\href {\doibase 10.1038/s41598-020-62020-z}
		{\bibfield  {journal} {\bibinfo  {journal} {Scientific reports}\ }\textbf
			{\bibinfo {volume} {10}},\ \bibinfo {pages} {6181} (\bibinfo {year}
			{2020})}\BibitemShut {NoStop}%
		\bibitem [{\citenamefont {Reichert}\ \emph
			{et~al.}(2018{\natexlab{a}})\citenamefont {Reichert}, \citenamefont
			{Defienne},\ and\ \citenamefont {Fleischer}}]{reichert_massively_2018}%
		\BibitemOpen
		\bibfield  {author} {\bibinfo {author} {\bibfnamefont {M.}~\bibnamefont
				{Reichert}}, \bibinfo {author} {\bibfnamefont {H.}~\bibnamefont {Defienne}},
			\ and\ \bibinfo {author} {\bibfnamefont {J.~W.}\ \bibnamefont {Fleischer}},\
		}\href@noop {} {\bibfield  {journal} {\bibinfo  {journal} {Scientific
					Reports}\ }\textbf {\bibinfo {volume} {8}},\ \bibinfo {pages} {7925}
			(\bibinfo {year} {2018})}\BibitemShut {NoStop}%
		\bibitem [{\citenamefont {Einstein}\ \emph {et~al.}(1935)\citenamefont
			{Einstein}, \citenamefont {Podolsky},\ and\ \citenamefont
			{Rosen}}]{Einstein1935}%
		\BibitemOpen
		\bibfield  {author} {\bibinfo {author} {\bibfnamefont {A.}~\bibnamefont
				{Einstein}}, \bibinfo {author} {\bibfnamefont {B.}~\bibnamefont {Podolsky}},
			\ and\ \bibinfo {author} {\bibfnamefont {N.}~\bibnamefont {Rosen}},\ }\href
		{\doibase 10.1103/PhysRev.47.777} {\bibfield  {journal} {\bibinfo  {journal}
				{Physical Review}\ }\textbf {\bibinfo {volume} {47}},\ \bibinfo {pages} {777}
			(\bibinfo {year} {1935})}\BibitemShut {NoStop}%
		\bibitem [{\citenamefont {Howell}\ \emph
			{et~al.}(2004{\natexlab{b}})\citenamefont {Howell}, \citenamefont {Bennink},
			\citenamefont {Bentley},\ and\ \citenamefont
			{Boyd}}]{howell_realization_2004}%
		\BibitemOpen
		\bibfield  {author} {\bibinfo {author} {\bibfnamefont {J.~C.}\ \bibnamefont
				{Howell}}, \bibinfo {author} {\bibfnamefont {R.~S.}\ \bibnamefont {Bennink}},
			\bibinfo {author} {\bibfnamefont {S.~J.}\ \bibnamefont {Bentley}}, \ and\
			\bibinfo {author} {\bibfnamefont {R.~W.}\ \bibnamefont {Boyd}},\ }\href
		{\doibase 10.1103/PhysRevLett.92.210403} {\bibfield  {journal} {\bibinfo
				{journal} {Physical Review Letters}\ }\textbf {\bibinfo {volume} {92}},\
			\bibinfo {pages} {210403} (\bibinfo {year} {2004}{\natexlab{b}})}\BibitemShut
		{NoStop}%
		\bibitem [{\citenamefont {Bavaresco}\ \emph
			{et~al.}(20018{\natexlab{b}})\citenamefont {Bavaresco}, \citenamefont {V	lencia},
			\citenamefont {Klöckl}, \citenamefont {Pivoluska}, \citenamefont {Erker}, \citenamefont {Friis}, \citenamefont {Malik},\ and\ \citenamefont
			{Herbert}}]{Bavaresco2018}%
		\BibitemOpen
		\bibfield  {author} {\bibinfo {author} {\bibfnamefont {J.~C.}\ \bibnamefont
				{Bavaresco}}, \bibinfo {author} {\bibfnamefont {N.~H.}\ \bibnamefont {Valencia}}, \bibinfo {author} {\bibfnamefont {C.}\ \bibnamefont {Klöckl}}
				, \bibinfo {author} {\bibfnamefont {P.}\ \bibnamefont {Pivoluska}}, \bibinfo {author} {\bibfnamefont {P.}\ \bibnamefont {Erker}},
			\bibinfo {author} {\bibfnamefont {N.}\ \bibnamefont {Friis}},
			\bibinfo {author} {\bibfnamefont {M.}\ \bibnamefont {Malik}}, \ and\
			\bibinfo {author} {\bibfnamefont {M.}\ \bibnamefont {Herbert}},\ }\href
		{\doibase 10.1038/s41567-018-0203-z} {\bibfield  {journal} {\bibinfo
				{journal} {Nature Physics}\ }\textbf {\bibinfo {number} {10}},\
			\bibinfo {pages} {1032} (\bibinfo {year} {2018}{\natexlab{b}})}\BibitemShut
		{NoStop}%
		\bibitem [{\citenamefont {Gasparini}\ \emph
			{et~al.}(2028{\natexlab{b}})\citenamefont {Gasparini}, \citenamefont {Zarghami}, 
			\citenamefont {Xu}, \citenamefont {Parmesan}, \citenamefont {Garcia}, \citenamefont {Unternährer}, \citenamefont {Bessire}, \citenamefont {Stefanov}, \citenamefont {Stoppa}, and\ \citenamefont
			{Perenzoni}}]{Gasparini2018}%
		\BibitemOpen
		\bibfield  {author} {\bibinfo {author} {\bibfnamefont {L.}\ \bibnamefont
				{Gasparini}}, \bibinfo {author} {\bibfnamefont {M.}\ \bibnamefont {Zarghami}},
			\bibinfo {author} {\bibfnamefont {H.}\ \bibnamefont {Xu}}, \bibinfo {author} {\bibfnamefont {L.}\ \bibnamefont
				{Parmesan}}, \bibinfo {author} {\bibfnamefont {M.~M.}\ \bibnamefont {Garcia}},
			\bibinfo {author} {\bibfnamefont {M.}\ \bibnamefont {Unternährer}}, \bibinfo {author} {\bibfnamefont {B.}\ \bibnamefont
				{Bessire}}, \bibinfo {author} {\bibfnamefont {A.}\ \bibnamefont {Stefanov}},
			\bibinfo {author} {\bibfnamefont {D.}\ \bibnamefont {Stoppa}}, \ and\
			\bibinfo {author} {\bibfnamefont {M.}\ \bibnamefont {Perenzoni}},\ }\href
		{\doibase 10.1109/ISSCC.2018.8310202} {\bibfield  {journal} {\bibinfo
				{journal} {2018 IEEE International Solid-State Circuits Conference - (ISSCC)}}\textbf {\bibinfo {volume} {}},\
			\bibinfo {pages} {98} (\bibinfo {year} {2018}{\natexlab{b}})}\BibitemShut
		{NoStop}%
		\bibitem [{\citenamefont {Moreau}\ \emph {et~al.}(2012)\citenamefont {Moreau},
			\citenamefont {Mougin-Sisini}, \citenamefont {Devaux},\ and\ \citenamefont
			{Lantz}}]{moreau_realization_2012}%
		\BibitemOpen
		\bibfield  {author} {\bibinfo {author} {\bibfnamefont {P.-A.}\ \bibnamefont
				{Moreau}}, \bibinfo {author} {\bibfnamefont {J.}~\bibnamefont
				{Mougin-Sisini}}, \bibinfo {author} {\bibfnamefont {F.}~\bibnamefont
				{Devaux}}, \ and\ \bibinfo {author} {\bibfnamefont {E.}~\bibnamefont
				{Lantz}},\ }\href {\doibase 10.1103/PhysRevA.86.010101} {\bibfield  {journal}
			{\bibinfo  {journal} {Physical Review A}\ }\textbf {\bibinfo {volume} {86}},\
			\bibinfo {pages} {010101} (\bibinfo {year} {2012})}\BibitemShut {NoStop}%
		\bibitem [{\citenamefont {Edgar}\ \emph {et~al.}(2012)\citenamefont {Edgar},
			\citenamefont {Tasca}, \citenamefont {Izdebski}, \citenamefont {Warburton},
			\citenamefont {Leach}, \citenamefont {Agnew}, \citenamefont {Buller},
			\citenamefont {Boyd},\ and\ \citenamefont {Padgett}}]{Edgar2012}%
		\BibitemOpen
		\bibfield  {author} {\bibinfo {author} {\bibfnamefont {M.}~\bibnamefont
				{Edgar}}, \bibinfo {author} {\bibfnamefont {D.}~\bibnamefont {Tasca}},
			\bibinfo {author} {\bibfnamefont {F.}~\bibnamefont {Izdebski}}, \bibinfo
			{author} {\bibfnamefont {R.}~\bibnamefont {Warburton}}, \bibinfo {author}
			{\bibfnamefont {J.}~\bibnamefont {Leach}}, \bibinfo {author} {\bibfnamefont
				{M.}~\bibnamefont {Agnew}}, \bibinfo {author} {\bibfnamefont
				{G.}~\bibnamefont {Buller}}, \bibinfo {author} {\bibfnamefont
				{R.}~\bibnamefont {Boyd}}, \ and\ \bibinfo {author} {\bibfnamefont
				{M.}~\bibnamefont {Padgett}},\ }\href {\doibase 10.1038/ncomms1988}
		{\bibfield  {journal} {\bibinfo  {journal} {Nature Communications}\ }\textbf
			{\bibinfo {volume} {3}},\ \bibinfo {pages} {984} (\bibinfo {year}
			{2012})}\BibitemShut {NoStop}%
		\bibitem [{\citenamefont {Defienne}\ \emph
			{et~al.}(2018{\natexlab{a}})\citenamefont {Defienne}, \citenamefont
			{Reichert},\ and\ \citenamefont {Fleischer}}]{defienne_general_2018-2}%
		\BibitemOpen
		\bibfield  {author} {\bibinfo {author} {\bibfnamefont {H.}~\bibnamefont
				{Defienne}}, \bibinfo {author} {\bibfnamefont {M.}~\bibnamefont {Reichert}},
			\ and\ \bibinfo {author} {\bibfnamefont {J.~W.}\ \bibnamefont {Fleischer}},\
		}\href@noop {} {\bibfield  {journal} {\bibinfo  {journal} {Physical Review
					Letters}\ }\textbf {\bibinfo {volume} {120}},\ \bibinfo {pages} {203604}
			(\bibinfo {year} {2018}{\natexlab{a}})}\BibitemShut {NoStop}%
		\bibitem [{\citenamefont {Fedorov}\ \emph {et~al.}(2009)\citenamefont
			{Fedorov}, \citenamefont
			{Mikhailova}, and\ \citenamefont
			{Volkov}}]{Fedorov2009}%
		\BibitemOpen
		\bibfield  {author} {\bibinfo {author} {\bibfnamefont {M.V.}~\bibnamefont
				{Fedorov}},\bibinfo {author} {\bibfnamefont {Yu.M.}~\bibnamefont
				{Mikhailova}} and \bibinfo {author} {\bibfnamefont {P.A.}~\bibnamefont {Volkov}}, }\href {\doibase
			10.1088/0953-4075/42/17/175503} {\bibfield  {journal} {\bibinfo  {journal} {Journal of Physics B} \textbf {\bibinfo {volume} {17}},\ \bibinfo
			{pages} {175503} (\bibinfo {year} {2009})}}\ \BibitemShut {NoStop}%
		\bibitem [{\citenamefont {Schneeloch}\ \emph {et~al.}(2016)\citenamefont
			{Schneeloch}, and\ \citenamefont
			{Howell}}]{Schneeloch2016}%
		\BibitemOpen
		\bibfield  {author} {\bibinfo {author} {\bibfnamefont {J.}~\bibnamefont
				{Schneeloch}} and \bibinfo {author} {\bibfnamefont {J.C.}~\bibnamefont {Howell}}, }\href {\doibase
			10.1088/2040-8978/18/5/053501} {\bibfield  {journal} {\bibinfo  {journal} {Journal of Optics} \textbf {\bibinfo {volume} {5}},\ \bibinfo
			{pages} {053501} (\bibinfo {year} {2016})}}\ \BibitemShut {NoStop}%
		\bibitem [{\citenamefont {Morimoto}\ \emph
			{et~al.}(2018{\natexlab{a}})\citenamefont {Morimoto}, \citenamefont
			{Ardelean}, \citenamefont{Wu}, \citenamefont{Ulku}, \citenamefont{Antolovic}, \citenamefont{Bruschini},\ and\ \citenamefont {Charbon}}]{Morimoto2020_1}%
		\BibitemOpen
		\bibfield  {author} {\bibinfo {author} {\bibfnamefont {K.}~\bibnamefont
				{Morimoto}}, \bibinfo {author} {\bibfnamefont {A.}~\bibnamefont {Ardelean}}, \bibinfo {author} {\bibfnamefont {M.-L.}~\bibnamefont {Wu}}, \bibinfo {author} {\bibfnamefont {A.~C.}~\bibnamefont {Ulku}}, \bibinfo {author} {\bibfnamefont {I.~M}~\bibnamefont {Antolovic}}, \bibinfo {author} {\bibfnamefont {C.}~\bibnamefont {Bruschini}},
			\ and\ \bibinfo {author} {\bibfnamefont {E.}\ \bibnamefont {Charbon}},\
		}\href{10.1364/OPTICA.386574} {\bibfield  {journal} {\bibinfo  {journal} {Optica}\ }\textbf {\bibinfo {volume} {7}},\ \bibinfo {pages} {346}
			(\bibinfo {year} {2020}{\natexlab{a}})}\BibitemShut {NoStop}%
		\bibitem [{\citenamefont {Morimoto}\ \emph
			{et~al.}(2018{\natexlab{a}})\citenamefont {Morimoto},\ and\ \citenamefont {Charbon}}]{Morimoto2020_2}%
		\BibitemOpen
		\bibfield  {author} {\bibinfo {author} {\bibfnamefont {K.}~\bibnamefont
				{Morimoto}},\ and\ \bibinfo {author} {\bibfnamefont {E.}\ \bibnamefont {Charbon}},\
		}\href{10.1364/OE.389216} {\bibfield  {journal} {\bibinfo  {journal} {Optics Express}\ }\textbf {\bibinfo {volume} {9}},\ \bibinfo {pages} {13068}
			(\bibinfo {year} {2020}{\natexlab{a}})}\BibitemShut {NoStop}%
		\bibitem [{\citenamefont {Reichert}\ \emph
			{et~al.}(2018{\natexlab{b}})\citenamefont {Reichert}, \citenamefont
			{Defienne},\ and\ \citenamefont {Fleischer}}]{reichert_optimizing_2018-3}%
		\BibitemOpen
		\bibfield  {author} {\bibinfo {author} {\bibfnamefont {M.}~\bibnamefont
				{Reichert}}, \bibinfo {author} {\bibfnamefont {H.}~\bibnamefont {Defienne}},
			\ and\ \bibinfo {author} {\bibfnamefont {J.~W.}\ \bibnamefont {Fleischer}},\
		}\href {\doibase 10.1103/PhysRevA.98.013841} {\bibfield  {journal} {\bibinfo
				{journal} {Physical Review A}\ }\textbf {\bibinfo {volume} {98}},\ \bibinfo
			{pages} {013841} (\bibinfo {year} {2018}{\natexlab{b}})}\BibitemShut
		{NoStop}%
		\bibitem [{\citenamefont {Erker}\ \emph
			{et~al.}(2017{\natexlab{b}})\citenamefont {Erker}, \citenamefont
			{Krenn},\ and\ \citenamefont {Hubert}}]{Erker2017}%
		\BibitemOpen
		\bibfield  {author} {\bibinfo {author} {\bibfnamefont {P.}~\bibnamefont
				{Erker}}, \bibinfo {author} {\bibfnamefont {M.}~\bibnamefont {Krenn}},
			\ and\ \bibinfo {author} {\bibfnamefont {M.}\ \bibnamefont {Huber}},\
		}\href {\doibase 10.22331/q-2017-07-28-22} {\bibfield  {journal} {\bibinfo
				{journal} {Quantum}\ }\textbf {\bibinfo {volume} {1}},\ \bibinfo
			{pages} {22} (\bibinfo {year} {2017}{\natexlab{b}})}\BibitemShut
		{NoStop}%
		\bibitem [{\citenamefont {Valencia}\ \emph
			{et~al.}(2020{\natexlab{b}})\citenamefont {Valencia}, \citenamefont
			{Srivastav}, \citenamefont {Pivoluska}, \citenamefont {Huber}, \citenamefont {Friis}, \citenamefont {McCutcheon},\ and\ \citenamefont {Malik}}]{Valencia2020}%
		\BibitemOpen
		\bibfield  {author} {\bibinfo {author} {\bibfnamefont {N.H.}~\bibnamefont
				{Valencia}}, \bibinfo {author} {\bibfnamefont {V.}~\bibnamefont {Srivastav}}, \bibinfo {author} {\bibfnamefont {M.}~\bibnamefont {Pivoluska}}, \bibinfo {author} {\bibfnamefont {M.}~\bibnamefont {Hubert}},
				 \bibinfo {author} {\bibfnamefont {N.}~\bibnamefont {Friis}}, \bibinfo {author} {\bibfnamefont {W.}~\bibnamefont {McCutcheon}},
			\ and\ \bibinfo {author} {\bibfnamefont {M.}\ \bibnamefont {Malik}},\ }\href
		{http://arxiv.org/abs/2004.04994} {\bibfield  {journal} {\bibinfo  {journal}
				{arXiv:2004.04994}\ } (\bibinfo {year}{2020}{\natexlab{b}})}\ \BibitemShut
		{NoStop}%
		\bibitem [{\citenamefont {Lantz}\ \emph {et~al.}(2015)\citenamefont {Lantz},
			\citenamefont {Denis}, \citenamefont {Moreau},\ and\ \citenamefont
			{Devaux}}]{lantz_einstein-podolsky-rosen_2015}%
		\BibitemOpen
		\bibfield  {author} {\bibinfo {author} {\bibfnamefont {E.}~\bibnamefont
				{Lantz}}, \bibinfo {author} {\bibfnamefont {S.}~\bibnamefont {Denis}},
			\bibinfo {author} {\bibfnamefont {P.-A.}\ \bibnamefont {Moreau}}, \ and\
			\bibinfo {author} {\bibfnamefont {F.}~\bibnamefont {Devaux}},\ }\href
		{\doibase 10.1364/OE.23.026472} {\bibfield  {journal} {\bibinfo  {journal}
				{Optics Express}\ }\textbf {\bibinfo {volume} {23}},\ \bibinfo {pages}
			{26472} (\bibinfo {year} {2015})}\BibitemShut {NoStop}%
			\bibitem [{\citenamefont {Defienne}\ \emph
			{et~al.}(2018{\natexlab{b}})\citenamefont {Defienne}, \citenamefont
			{Reichert},\ and\ \citenamefont {Fleischer}}]{defienne_adaptive_2018-1}%
		\BibitemOpen
		\bibfield  {author} {\bibinfo {author} {\bibfnamefont {H.}~\bibnamefont
				{Defienne}}, \bibinfo {author} {\bibfnamefont {M.}~\bibnamefont {Reichert}},
			\ and\ \bibinfo {author} {\bibfnamefont {J.~W.}\ \bibnamefont {Fleischer}},\
		}\href {\doibase 10.1103/PhysRevLett.121.233601} {\bibfield  {journal}
			{\bibinfo  {journal} {Physical Review Letters}\ }\textbf {\bibinfo {volume}
				{121}},\ \bibinfo {pages} {233601} (\bibinfo {year}
			{2018}{\natexlab{b}})}\BibitemShut {NoStop}%
		\bibitem [{\citenamefont {Defienne}\ \emph
			{et~al.}(2019{\natexlab{b}})\citenamefont {Defienne}, \citenamefont
			{Ndagano}, \citenamefont {Lyons},\ and\ \citenamefont
			{Faccio}}]{defienne_entanglement-enabled_2019}%
		\BibitemOpen
		\bibfield  {author} {\bibinfo {author} {\bibfnamefont {H.}~\bibnamefont
				{Defienne}}, \bibinfo {author} {\bibfnamefont {B.}~\bibnamefont {Ndagano}},
			\bibinfo {author} {\bibfnamefont {A.}~\bibnamefont {Lyons}}, \ and\ \bibinfo
			{author} {\bibfnamefont {D.}~\bibnamefont {Faccio}},\ }\href
		{http://arxiv.org/abs/1911.01209} {\bibfield  {journal} {\bibinfo  {journal}
				{arXiv:1911.01209}\ } (\bibinfo {year}{2019}{\natexlab{b}})}\ \BibitemShut
		{NoStop}%
		\bibitem [{\citenamefont {Schneeloch}\ \emph
			{et~al.}(2019{\natexlab{a}})\citenamefont {Tison}, \citenamefont
			{Fanto}, \citenamefont {Alsing},\ and\ \citenamefont
			{Howland}}]{Schneeloch2019}%
		\BibitemOpen
		\bibfield  {author} {\bibinfo {author} {\bibfnamefont {J.}~\bibnamefont
				{Schneeloch}}, \bibinfo {author} {\bibfnamefont {C.}~\bibnamefont
				{Tison}}, \bibinfo {author} {\bibfnamefont {M.}~\bibnamefont
				{Fanto}}, \bibinfo {author} {\bibfnamefont {P.M.}~\bibnamefont {Alsing}}, \ and\ \bibinfo {author} {\bibfnamefont
				{G.A.}~\bibnamefont {Howland}},\ }\href {\doibase 10.1038/s41467-019-10810-z} {\bibfield
			{journal} {\bibinfo  {journal} {Nature Communications}\ }\textbf {\bibinfo
				{volume} {10}},\ \bibinfo {pages} {2785} (\bibinfo {year}
			{2019}{\natexlab{a}})}\BibitemShut {NoStop}%
		\bibitem [{\citenamefont {Valencia}\ \emph
			{et~al.}(2020{\natexlab{b}})\citenamefont {Valencia}, \citenamefont {Goel}, \citenamefont {McCutcheon}, \citenamefont {Defienne},\ and\ \citenamefont {Malik}}]{Valencia2019}%
		\BibitemOpen
		\bibfield  {author} {\bibinfo {author} {\bibfnamefont {N.H.}~\bibnamefont
				{Valencia}}, \bibinfo {author} {\bibfnamefont {S.}~\bibnamefont {Goel}}, \bibinfo {author} {\bibfnamefont {W.}~\bibnamefont {McCutcheon}}, \bibinfo {author} {\bibfnamefont {H.}~\bibnamefont {Defienne}},
			\ and\ \bibinfo {author} {\bibfnamefont {M.}\ \bibnamefont {Malik}},\ }\href
		{http://arxiv.org/abs/1910.04490} {\bibfield  {journal} {\bibinfo  {journal}
				{arXiv:1910.04490}\ } (\bibinfo {year}{2019}{\natexlab{b}})}\ \BibitemShut
		{NoStop}%
		\bibitem [{\citenamefont {Fontaine}\ \emph
			{et~al.}(2019{\natexlab{a}})\citenamefont {Ryf}, \citenamefont
			{Chen}, \citenamefont {Neilson}, \citenamefont {Kim},\ and\ \citenamefont
			{Carpenter}}]{Fontaine2019}%
		\BibitemOpen
		\bibfield  {author} {\bibinfo {author} {\bibfnamefont {N.K.}~\bibnamefont
				{Fontaine}}, \bibinfo {author} {\bibfnamefont {R.}~\bibnamefont
				{Ryf}}, \bibinfo {author} {\bibfnamefont {H.}~\bibnamefont
				{Chen}}, \bibinfo {author} {\bibfnamefont {D.T.}~\bibnamefont {Neilson}},
			\bibinfo {author} {\bibfnamefont {K.}\ \bibnamefont {Kim}}, \ and\ \bibinfo {author} {\bibfnamefont
				{J.}~\bibnamefont {Carpenter}},\ }\href {\doibase 10.1038/s41467-019-09840-4} {\bibfield
			{journal} {\bibinfo  {journal} {Nature Communications}\ }\textbf {\bibinfo
				{volume} {10}},\ \bibinfo {pages} {1865} (\bibinfo {year}
			{2019}{\natexlab{a}})}\BibitemShut {NoStop}%
		\bibitem [{\citenamefont {Brandt}\ \emph
			{et~al.}(2020{\natexlab{a}})\citenamefont {Hiekkamäki}, \citenamefont
			{Bouchard}, \citenamefont{Hubert},\ and\ \citenamefont {Fickler}}]{Brandt2020}%
		\BibitemOpen
		\bibfield  {author} {\bibinfo {author} {\bibfnamefont {F.}~\bibnamefont
				{Brandt}}, \bibinfo {author} {\bibfnamefont {M.}~\bibnamefont {Hiekkamäki}}, \bibinfo {author} {\bibfnamefont {F.}~\bibnamefont {Bouchard}}, \bibinfo {author} {\bibfnamefont {M.}~\bibnamefont {Hubert}}, ,
			\ and\ \bibinfo {author} {\bibfnamefont {R.}\ \bibnamefont {Fickler}},\
		}\href{10.1364/OPTICA.3758754} {\bibfield  {journal} {\bibinfo  {journal} {Optica}\ }\textbf {\bibinfo {volume} {7}},\ \bibinfo {pages} {98}
			(\bibinfo {year} {2020}{\natexlab{a}})}\BibitemShut {NoStop}%
		\bibitem [{\citenamefont {Larchuk}\ \emph {et~al.}(1995)\citenamefont
			{Larchuk}, \citenamefont {Teich},\ and\ \citenamefont
			{Saleh}}]{larchuk_statistics_1995}%
		\BibitemOpen
		\bibfield  {author} {\bibinfo {author} {\bibfnamefont {T.~S.}\ \bibnamefont
				{Larchuk}}, \bibinfo {author} {\bibfnamefont {M.~C.}\ \bibnamefont {Teich}},
			\ and\ \bibinfo {author} {\bibfnamefont {B.~E.~A.}\ \bibnamefont {Saleh}},\
		}\href {\doibase 10.1111/j.1749-6632.1995.tb39009.x} {\bibfield  {journal}
			{\bibinfo  {journal} {Annals of the New York Academy of Sciences}\ }\textbf
			{\bibinfo {volume} {755}},\ \bibinfo {pages} {680} (\bibinfo {year}
			{1995})}\BibitemShut {NoStop}%
	\end{thebibliography}

\smallskip
\noindent * These authors contributed equally to this work. \\
\textsuperscript{\textdagger} email: bienvenu.ndagano@glasgow.ac.uk \\
\textsuperscript{$\ddagger$} email: daniele.faccio@glasgow.ac.uk 

\section{Acknowledgements} 
The authors thanks M. Malik and N.H. Valencia for fruitful discussions. D.F. acknowledges financial support from the UK Engineering and Physical Sciences Research Council (grants EP/M01326X/1 and EP/R030081/1) and from the European Union's Horizon 2020 research and innovation programme under grant agreement No 801060. H.D. acknowledges financial support from the EU Marie-Curie Sklodowska Actions (project \textsl{840958}).

\section{Authors contributions} 
B.N. performed the experiment. B.N. and H.D. analysed the results. B.N., H.D., A.L. and D.F. conceived and discussed the experiment. F.V. and S.T. provided experimental support with the SPAD camera. All authors contributed to the redaction of the manuscript. 

\section{Competing interests} 
The authors declare no competing interests. \\

\BN{\section{Figure legends}
Figure~\ref{Figure1}: \textbf{Experimental scheme.} Spatially-entangled photon pairs are produced by spontaneous parametric down-conversion (SPDC) in a $\beta$-barium borate (BBO) using a $347$~nm pulsed pump laser with a repetition rate of 100~MHz. Spectral filters (SF) select near-degenerate photon pairs at $694 \pm 5$~nm. \textbf{a,} A three-lens system composed of $f_1=35$~mm, $f_2=100$~mm and $f_3=200$~mm, maps photons momenta onto pixels of a single-photon avalanche diode (SPAD) camera by Fourier imaging the crystal. This configuration is named FF-configuration.  \textbf{b,} To measure position correlations, two lenses, $f_1=35$~mm and $f_4=300$~mm, image the output surface of the BBO crystal onto the SPAD camera. This configuration is named NF-configuration.

Figure~\ref{Figure2}: \textbf{Measurement of momentum correlations.} \textbf{a,} Intensity distribution of the SPDC light measured in the FF-configuration. \textbf{b} and \textbf{c}, Conditional probability distributions $\Gamma(\vec{k}|\vec{A})$ and  $\Gamma(\vec{k}|\vec{B})$ relative to two arbitrarily chosen positions $\vec{A}$ and $\vec{B}$ on the sensor, respectively. We measured an SNR of 320 (\textbf{b}) and 258 (\textbf{c}). \textbf{d}, Projection of the joint probability distribution (JPD) along the sum coordinates $\vec{k_1}+\vec{k_2}$. A measured momentum correlation width of $\Delta\vec{k} = 1.0666(7)\times 10^{-3} $ rad.$\mu\text{m}^{-1}$ is obtained using a Gaussian fit (see Methods). Spatial coordinates are in pixels and the analysis was performed on a total of $10^7$ images.

Figure~\ref{Figure3}: \textbf{Measurement of position correlations.} \textbf{a,} Intensity distribution of the SPDC light in the NF-configuration. \textbf{b,} Projection of joint probability distribution along the minus-coordinates $\vec{r_1}-\vec{r_2}$. A measured position correlation width $\Delta\vec{r} = 4.3(2)$ $\mu\text{m}$ was obtained using a Gaussian fit (see Methods). Spatial coordinates are in pixels and the analysis was performed on a total of $10^7$ images.

Figure~\ref{Figure4}: \textbf{Confidence level analysis}. \textbf{a,} $C$ values measured for different total number of frames $N$ (black crosses) together with a theoretical model of the form $0.047 \sqrt{N}$ (blue dashed line). \textbf{b-d}, Sum-coordinate projections of the JPD measured in the FF-configuration using \textbf{(b)} $2 \times 10^3$ frames, \textbf{(c)} $10^4$ frames and \textbf{(d)} $10^7$ frames. \textbf{e-g}, Minus-coordinate projection of the JPD measured in the NF configuration using \textbf{(e)} $2 \times 10^3$ frames, \textbf{(f)} $10^4$ frames and \textbf{(g)} $10^7$ frames.

Figure~\ref{Figure5}:\textbf{Entanglement certification measurements}. In \textbf{a} near-field  and \textbf{b} far-field configurations, we select the same square grid of 14$\times$14 = 196 pixels, with a spacing of 1 pixel (grid of black squares). \textbf{c} and \textbf{d} show the measured correlations between all the pixels in the grid, with each pixel labelling the spatial coordinates of photon: $\ket{m}$ in the near-field (position) and $\ket{\tilde{p}}$ in the far-field (momentum). We calculated a lower bound $\tilde{F} = 0.252(9)$ of the fidelity with respect to a $d$~=~196 maximally entangled state, leading to a certified entanglement dimensionality of 48. Spatial coordinates are in pixels and the analysis was performed on a total of $10^7$ images.}

\end{document}